\documentclass[aps,amsmath,prd,showpacs,nofootinbib,amssymb,preprint]{revtex4}
\pdfoutput=1 

\usepackage{amssymb,graphics,graphicx,epstopdf,color,subfigure,dcolumn,bm,slashed,tabularx,bm,amssymb}

\def\roughly#1{\mathrel{\raise.3ex\hbox
{$#1$\kern-.75em\lower1ex\hbox{$\sim$}}}}

\begin{document}

\title{Probing Dark Matter Self-Interaction in the Sun with IceCube-PINGU}
\author{Chian-Shu~Chen$^{1,3}$\footnote{chianshu@phys.sinica.edu.tw}, Fei-Fan Lee$^{2}$\footnote{fflee@mail.nctu.edu.tw}, Guey-Lin Lin$^{2}$\footnote{glin@cc.nctu.edu.tw}, and Yen-Hsun Lin$^{2}$\footnote{chris.py99g@g2.nctu.edu.tw}}
  \affiliation{$^{1}$Physics Division, National Center for Theoretical Sciences, Hsinchu 30010, Taiwan\\
$^{2}$Institute of Physics, National Chiao Tung University Hsinchu 30010, Taiwan\\
$^{3}$Department of Physics, National Tsing Hua University, Hsinchu 30010, Taiwan}

\date{\today}
\begin{abstract}
We study the capture, annihilation and evaporation of dark matter (DM) inside the Sun. It has been shown that the DM self-interaction can increase the DM number inside the Sun.  
We demonstrate that this enhancement becomes more significant  in the regime of small DM mass, given a fixed DM self-interaction cross section. 
This leads to the enhancement of neutrino flux from DM annihilation.
On the other hand, for DM mass as low as a few GeVs, not only the DM-nuclei scatterings can cause the DM evaporation, DM self-interaction also provides 
non-negligible contributions to this effect. Consequently, the critical mass for DM evaporation (typically 3 $\sim$ 4 GeV without the DM self-interaction) 
can be slightly  increased.  We discuss the prospect of detecting DM self-interaction in IceCube-PINGU using the annihilation 
channels $\chi\chi \rightarrow \tau^{+}\tau^{-}, \nu\bar{\nu}$ as examples. The PINGU sensitivities to DM self-interaction cross section $\sigma_{\chi\chi}$ are estimated for track and cascade events.   
\end{abstract}

\pacs{14.60.Pq, 12.60.-i, 14.80.-j, 14.80.Cp}

\maketitle

\section{Introduction}
Convincing observational evidences indicate that roughly 80\% of the matter in our universe is dark matter (DM). On the other hand, the understanding of the DM nature is still at the budding stage. If the form of DM is any kind of elementary particle $\chi$, theoretical predictions for its mass lie in a wide range, from sub-eV scale to grand unified scale.  Even though, more and more experimental searches, including direct and indirect detection with terrestrial or satellite instruments, are looking for the DM signals.  The knowledge on DM is accumulating rapidly. If DM interacts with ordinary matters weakly, it may leave tracks in the detector and be caught directly. The possibility to see such signals depends on DM mass and its interaction cross section with the ordinary matter in the detector. The situation is similar for the indirect searches where one looks for the flux excess in cosmic rays, gamma rays, and neutrinos.  The flux excess for the above particles are due to DM annihilation or DM decays. Different search strategies would have sensitivities to different parameter ranges and all together cover quite a broad range for DM mass and DM interaction cross section, respectively.  

In this paper, we study the general framework of DM capture, annihilation and evaporation in the Sun. 
The capture of galactic DM by the Sun through DM-nuclei collisions was first proposed and calculated in Refs.~\cite{Steigman:1997vs,Press:1985ug,Faulkner:1985rm,Griest:1986yu,Jungman:1995df,Bertone:2004pz}. It was then observed that the assumption of DM thermal distribution according to the average temperature of the Sun is a good approximation for capture and annihilation processes, but the correction to the evaporation mass can reach to 8\% in the true distribution calculation~\cite{Gould:1987ju}. 
The abundance of DM inside the Sun hence results from the balancing among DM capture, annihilation and  evaporation processes. Updated calculations on these processes are given in Refs.~\cite{Kappl:2011kz,Bernal:2012qh,Busoni:2013kaa}. 

In our study, we particularly note that for both collisionless cold DM and warm DM there exists 
a so-called core/cusp problem~\cite{deBlok:2009sp} which addresses the discrepancy between the 
computational structure simulation and the actual observation~\cite{Moore:1994yx,Flores:1994gz,Navarro:1996gj,Walker:2011zu,
Oh:2010ea,Maccio:2012qf}. DM self-interaction has been introduced to resolve this inconsistency~\cite{Spergel:1999mh}. 
This type of analyses put constraints on the ratio of DM self-interaction cross section to the DM mass, $0.1 < \sigma_{\chi\chi}/m_{\chi} < 1.0 ~ (\rm{cm}^2/g)$, from observations of various galactic structures~\cite{Randall:2007ph,Rocha:2012jg,Peter:2012jh,Zavala:2012us}. It is worth mentioning that the authors in Ref.~\cite{Albuquerque:2013xna} used the IceCube data~\cite{Aartsen:2012kia} to constrain the magnitude of $\sigma_{\chi\chi}$ for $m_{\chi}$ in the range of ${\cal O}(10)$~GeV to ${\cal O}(1)$~TeV (also see the study of high energy neutrino flux from DM annihilation within the Sun with the inclusion of DM self-interaction in Ref.~\cite{Zentner2009}).
In their work the evaporation effect can be neglected for the considered DM  mass region and the detectability of $\sigma_{\chi\chi}$ by the IceCube through observing the final state neutrino flux produced by the DM annihilation is analyzed. 
In this paper we shall concentrate on the low mass region of ${\cal O}(1)$~GeV DM mass since such a mass range has not been probed by the IceCube data mentioned 
above. Furthermore,  this is also the mass range where the indirect search is crucial.  In the case of spin-independent interaction, the sensitivity of DM direct search quickly turns poor for $m_{\chi}$ less than $10$ GeV \cite{Cushman:2013zza}. Therefore the IceCube-PINGU~\cite{Aartsen2014} detector with a $1$ GeV threshold energy could be more sensitive than some of the direct detection experiments for $m_{\chi}< 10$ GeV. For spin-dependent interaction, the IceCube-PINGU sensitivity has been estimated to be much better than constraints set by direct detection experiments~\cite{Feng:2014uja}.         

For illustrative purpose, we shall only consider neutrino flux produced by DM annihilating into leptons such as $\chi\chi\to \tau^{+}\tau^{-}$ and $\chi\chi\to \nu\bar{\nu}$, respectively. The final-state neutrinos can be detected 
by terrestrial neutrino telescopes such as IceCube-PINGU~\cite{Aartsen2014}.  We do not consider $\chi\chi\to \mu^+\mu^-$  because muons will suffer severe energy losses in the Sun before 
they decay to neutrinos. The soft neutrino spectrum in this case is dominated by the atmospheric background.  One also expects that the neutrino telescopes are less sensitive to heavy quark channels such as $\chi\chi\to b\bar{b}$ than their sensitivities 
to leptonic channels. This is caused by the relatively softer neutrino spectrum resulting from the $b$-hadron decays compared to the 
neutrino spectrum from $\tau$ decays~\cite{spectrum}. For light quark channels $\chi\chi\to q\bar{q}$, the hadronic cascades produce pions or kaons in 
large multiplicities. These hadrons decay almost at rest and produce MeV neutrino fluxes, which from the observational point of view could be as promising as the 
hard spectrum channels~\cite{Bernal:2012qh,Rott:2012qb}. However we would not discuss these signatures because the threshold energy of IceCube-PINGU 
is in the GeV range. 
   
For ${\cal O}(1)$~GeV DM mass, it is necessary to consider evaporation processes for determining the DM abundance inside the Sun. 
We shall demonstrate that DM evaporation can arise both from DM-nuclei scatterings and DM self-interactions. In fact 
the inclusion of DM self-interaction can  raise the critical mass such that evaporation can take place for DM lighter than this mass scale. On the other hand, DM self-interaction also enhances total DM number trapped inside 
the Sun for $m_{\chi}$ greater than the critical mass.

This paper is organized as follows. In Section II, we calculate the DM number accumulated inside the Sun via four processes.  We then present  the 
parameter space in which the equilibrium condition holds. In Section III, we present the numerical results of our analysis.
The sensitivity of IceCube-PINGU to DM self-interaction inside the Sun is presented  in Section IV. Finally, we conclude in Section V. 

\section{DM accumulation in the Sun}

\subsection{DM evolution equation\label{sec:WIMP_evo_eq}}
We assume that both the galactic DM and the nuclei inside the Sun follow the thermal distributions. If DM interacts with nuclei in the Sun then it can be captured by the Sun when its final velocity is smaller than the escape velocity from the Sun.
 Alternatively, DM trapped inside the Sun will be kicked out if its final velocity after the scattering with the nuclei is larger than the escape velocity. The inclusion of DM self-interaction will also have effects on 
the capture and evaporation of DM inside the Sun. 
We will come to details of  this in the later discussion.
 The captured DM could have reached to an equilibrium state if the equilibrium time scale is less than the age of the Sun ($t_{\odot} \approx 10^{17} s$). 
The general DM evolution equation in the Sun is given by
\begin{equation}
\frac{dN_{\chi}}{dt}=C_{c}+(C_{s}-C_{e})N_{\chi}-(C_{a}+C_{se}){N_{\chi}}^{2}\label{eq:evo_eq}
\end{equation}
with $N_{\chi}$ the DM number in the Sun,  $C_c$ the rate at which DM are captured by the Sun, $C_s$ the rate at which DM are captured due to their scattering with DM that have already been trapped in the Sun, $C_{e}$ the the DM evaporation rate due to DM-nuclei interactions, $C_{a}$ the DM annihilation rate,  and $C_{se}$ the evaporation rate induced by the interaction between DM particles in the Sun.
The coefficients $C_{a,c,e,s,se}$ are taken to be positive and \emph{time-independent}. 

We note that $C_c$ can be categorized by the type
of interactions between DM particles and nucleons. For spin-dependent (SD) interactions, the capture rate is given by~\cite{Jungman:1995df,Bertone:2004pz}
\begin{equation}
C_{c}^{\textrm{SD}}\simeq3.35\times10^{24}\textrm{ s}^{-1}\left(\frac{\rho_{0}}{0.3\textrm{ GeV/cm}^{3}}\right)\left(\frac{270\textrm{ km/s}}{\bar{v}}\right)^{3}\left(\frac{\textrm{GeV}}{m_{\chi}}\right)^{2}\left(\frac{\sigma_{\textrm{H}}^{\textrm{SD}}}{10^{-6}\textrm{ pb}}\right),\label{eq:capture_SD}
\end{equation}
where $\rho_{0}$ is the local DM density, $\bar{v}$ is the
velocity dispersion, $\sigma_{\textrm{H}}^{\textrm{SD}}$ is the SD
DM-hydrogen scattering cross section and $m_{\chi}$ is the DM mass. The capture rate from spin-independent (SI)  scattering is given by~\cite{Jungman:1995df,Bertone:2004pz}
\begin{equation}
C_{c}^{\textrm{SI}}\simeq1.24\times10^{24}\textrm{ s}^{-1}\left(\frac{\rho_{0}}{0.3\textrm{ GeV/cm}^{3}}\right)\left(\frac{270\textrm{ km/s}}{\bar{v}}\right)^{3}\left(\frac{\textrm{GeV}}{m_{\chi}}\right)^{2}\left(\frac{2.6\sigma_{\textrm{H}}^{\textrm{SI}}+0.175\sigma_{\textrm{He}}^{\textrm{SI}}}{10^{-6}\textrm{ pb}}\right).\label{eq:capture_SI}
\end{equation}
Here $\sigma_{\textrm{H}}^{\textrm{SI}}$ and $\sigma_{\textrm{He}}^{\textrm{SI}}$
are SI DM-hydrogen and -helium cross sections respectively.
Taking the approximation $m_{p}\approx m_{n}$,
the DM-nucleus cross section $\sigma_{i}$ is related to DM-nucleon
cross section $\sigma_{\chi p}$ by
\begin{equation}
\sigma_{i}^{\textrm{SD}}=A^{2}\left(\frac{m_{\chi}+m_{p}}{m_{\chi}+m_{A}}\right)^{2}
\frac{4(J_i+1)}{3 J_i}\left|\left\langle S_{p,i}\right\rangle +\left\langle S_{p,i}\right\rangle \right|^{2}
\sigma_{\chi p}^{\textrm{SD}}
\end{equation}
for SD interactions and
\begin{equation}
\sigma_{i}^{\textrm{SI}}=A^{2}\left(\frac{m_{A}}{m_{p}}\right)^{2}\left(\frac{m_{\chi}+m_{p}}{m_{\chi}+m_{A}}\right)^{2}\sigma_{\chi p}^{\textrm{SI}}
\end{equation}
for SI interactions, where $A$ is the atomic number, $m_{A}$
the mass of the nucleus, $J_i$ the total angular momentum of the nucleus and $\left\langle S_{p,i}\right\rangle$ and $\left\langle S_{n,i}\right\rangle$ the spin expectation values of proton and of neutron averaged over the entire nucleus~\cite{Ellis:1987sh}.

The DM evaporation rate in the Sun, $C_e$, has been well investigated
in Refs.~\cite{Griest:1986yu,Gould:1987ju}. The evaporation rate is usually
ignored in the DM evolution equation since
it happens for a very low DM mass, $m_{\chi}\lesssim3\textrm{ GeV}$.
A updated calculation in Ref.~\cite{Busoni:2013kaa} has shown that,
for $m_{\chi}/m_{A}>1$,
\begin{equation}
C_{e}\simeq\frac{8}{\pi^{3}}\sqrt{\frac{2m_{\chi}}{\pi T_{\chi}(\bar{r})}}\frac{v_{\textrm{esc}}^{2}(0)}{\bar{r}^{3}}\exp\left(-\frac{m_{\chi} v_{\textrm{esc}}^{2}(0)}
{2T_{\chi}(\bar{r})}\right)\Sigma_{\textrm{evap}},\label{eq:evap_rate}
\end{equation}
where $v_{\textrm{esc}}(0)$ is the escape velocity from
the core of the Sun, $T_{\chi}$ is the DM temperature in the Sun,
and $\bar{r}$ is average DM orbit radius which is the mean DM distance
from the solar center. The quantity $\Sigma_{\textrm{evap}}$ is the
sum of the scattering cross sections of all the nuclei within a radius
$r_{95\%}$, where the solar temperature has dropped to $95\%$ of
the DM temperature. 
Although the approximate form of $C_e$ can be obtained as the above equation, we shall adopt the exact form 
of $C_{e}$  given in Ref.~\cite{Gould:1987ju} for our subsequent numerical calculations.

As stated before, $C_{s}$ is the DM capture rate by scattering off the DM that have been captured inside the Sun. 
This kind of scattering may result in the target dark matter particles being ejected from the Sun upon recoil.
However, because the escape speed from the Sun is sufficiently large, 
the effect of target DM ejection by recoil is only a small correction to the simple solar capture estimate. Hence  the self-capture rate in the Sun can be approximated by~\cite{Zentner2009}
\begin{equation}
C_{s}=\sqrt{\frac{3}{2}}n_{\chi}\sigma_{\chi\chi} v_{\rm{esc}}(R_{\odot})\frac{v_{\rm{esc}}(R_{\odot})}{\overline{v}}\left\langle \widehat{\phi}_{\chi}\right\rangle  \frac{\rm{erf}(\eta)}{\eta},
\end{equation}
where $\left\langle \widehat{\phi}_{\chi}\right\rangle \simeq 5.1$~\cite{Gould:1992} is a dimensionless average solar potential experienced by the captured DM within the Sun, $n_{\chi}$ is the local number density of halo DM, $\sigma_{\chi\chi}$ is the elastic scattering cross section of DM with themselves, $v_{\rm{esc}}(R_{\odot})$ is the Sun's  escape velocity at the surface, and $\eta^{2}=3(v_{\odot}/\overline{v})^{2}/2$ is the square of a dimensionless velocity of the Sun through the Galactic halo with $v_{\odot}=220~\rm{km/s}$  Sun's velocity and $\overline{v}=270~\rm{km/s}$  the local velocity dispersion of DM in the halo. 

$C_{a}$ is the annihilation coefficient given by \cite{Griest:1986yu}
\begin{equation}
C_{a}\simeq\frac{\left\langle \sigma v \right\rangle V_{2}}{V_{1}^{2}},
\end{equation}
where
\begin{equation}
V_{j}\simeq6.5\times10^{28}\textrm{ cm}^{3}\left(\frac{10\textrm{ GeV}}{jm_{\chi}}\right)^{3/2}
\end{equation}
is the DM effective volume inside the Sun and $\langle \sigma v \rangle$ is the relative velocity averaged annihilation cross section.

$C_{se}$ is the self-interaction induced evaporation. Since the DM can interact among themselves, DM trapped in the solar core could scatter with other trapped DM and results in the evaporation. 
Essentially, one of the DM particles could have velocity greater than the escape velocity after the scattering.  This process involves two DM particles just like annihilation. We note that both processes lead to the DM dissipation in the Sun. 
On the other hand $C_{se}$ does not produce neutrino flux as $C_a$ does. Since the derivation of $C_{se}$ has not been given in the previous literature, we present 
some details of the derivation in Appendix \ref{sec:appendix}~\cite{Comment}.

With $N_{\chi}(0)=0$ as the initial condition, the general solution to Eq.~(\ref{eq:evo_eq})
is
\begin{equation}
N_{\chi}(t)=\frac{C_{c}\tanh(t/\tau_{A})}{\tau_{A}^{-1}-(C_{s}-C_{e})\tanh(t/\tau_{A})/2},
\end{equation}
with
\begin{equation}
\tau_{A}=\frac{1}{\sqrt{C_{c}(C_{a}+C_{se})+(C_{s}-C_{e})^{2}/4}}
\end{equation}
the time-scale for the DM number in the Sun to reach the equilibrium. If the equilibrium state is achieved, i.e., $\tanh(t/\tau_{A})\sim1$,
one has 
\begin{equation}
N_{\chi,\textrm{eq}}=\frac{C_{s}-C_{e}}{2(C_{a}+C_{se})}+\sqrt{\frac{(C_{s}-C_{e})^{2}}{4(C_{a}+C_{se})^{2}}+\frac{C_{c}}{C_{a}+C_{se}}}.
\end{equation}
The DM annihilation rate in the Sun's core is given by
\begin{equation}\label{annihilation}
\Gamma_{A}=\frac{C_{a}}{2}N_{\chi}^{2}.
\end{equation}
By setting $C_s = C_{se} = 0$, we can recover the results in Refs.~\cite{Griest:1986yu,Gould:1987ju,Kappl:2011kz,Bernal:2012qh,Busoni:2013kaa} for the absence of DM self-interaction. By setting $C_e = C_{se} = 0$, we recover the result  in Ref.~\cite{Zentner2009}, which includes the DM self-interaction while neglects 
the DM evaporation.

\subsection{Numerical results}

\begin{figure}[t]
\begin{centering}
\includegraphics[width=0.49\textwidth]{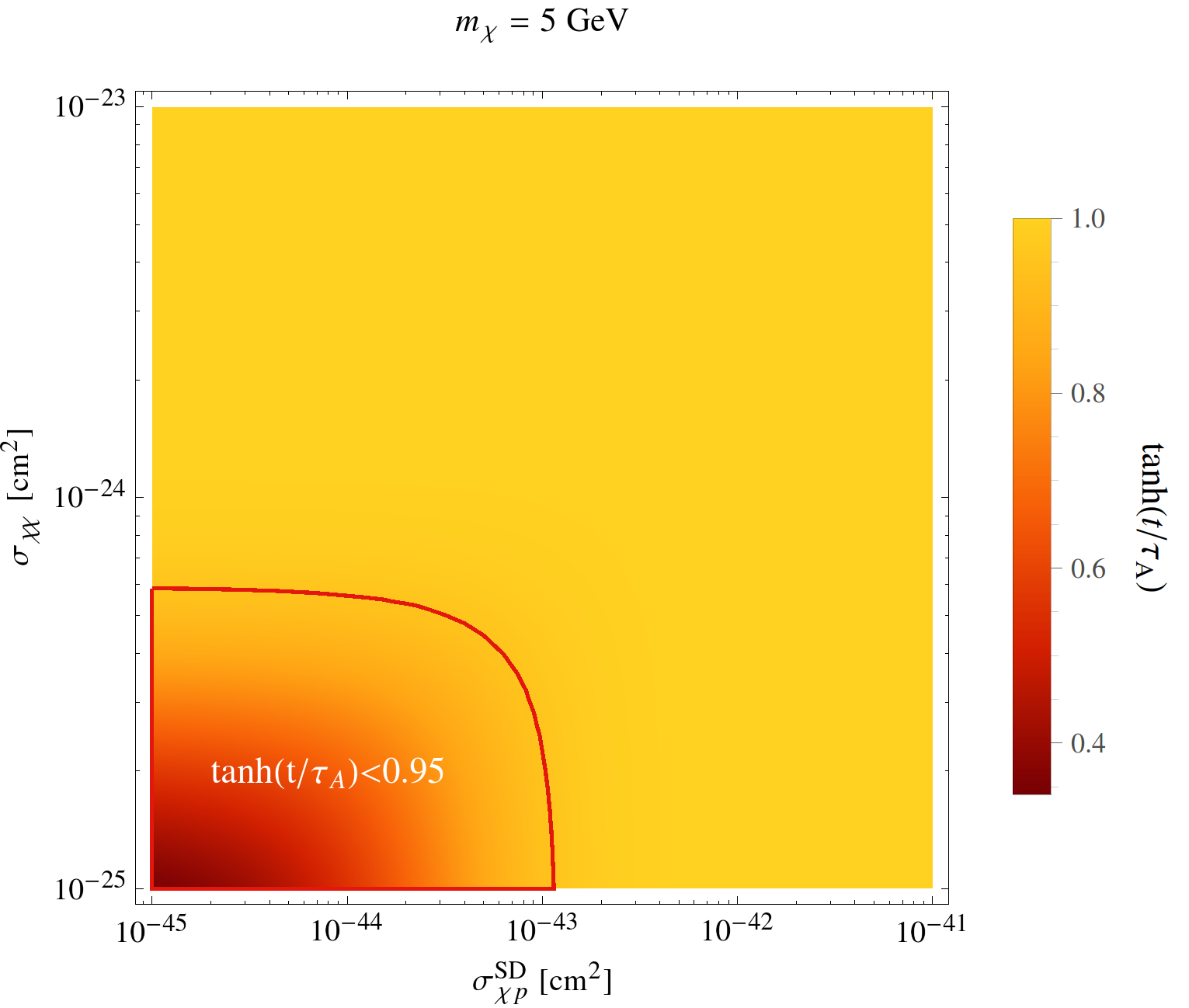}\includegraphics[width=0.49\textwidth]{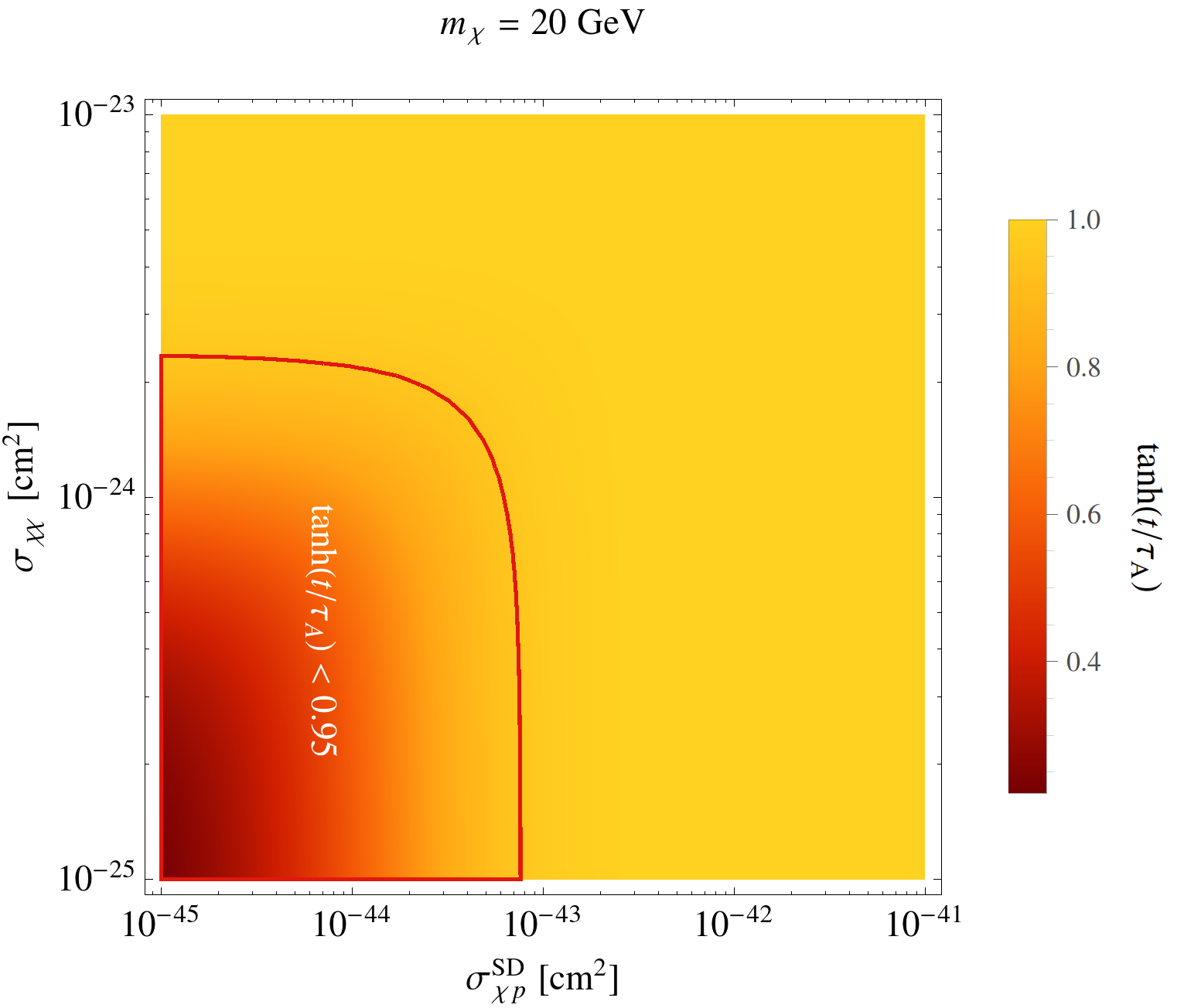}
\par\end{centering}
\caption{\label{fig:tanh_SD}The values of $\tanh(t/\tau_{A})$ over $\sigma_{\chi p}^{\textrm{SD}}-\sigma_{\chi\chi}$
plane at the present day, $t=t_{\odot}$. The red-circled area is the non-equilibrium region for $N_{\chi}$.}
\end{figure}
\begin{figure}[t]
\begin{centering}
\includegraphics[width=0.49\textwidth]{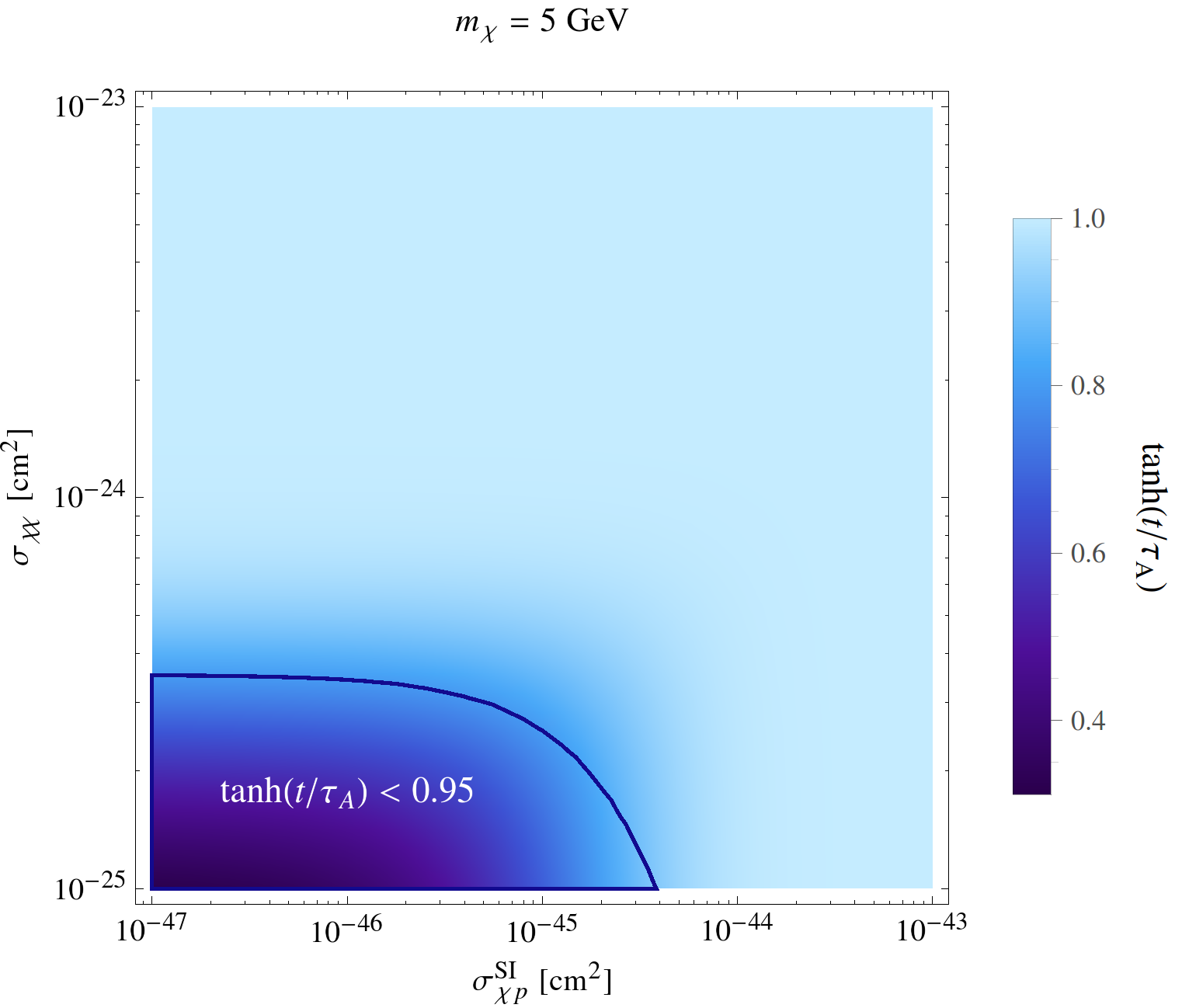}\includegraphics[width=0.49\textwidth]{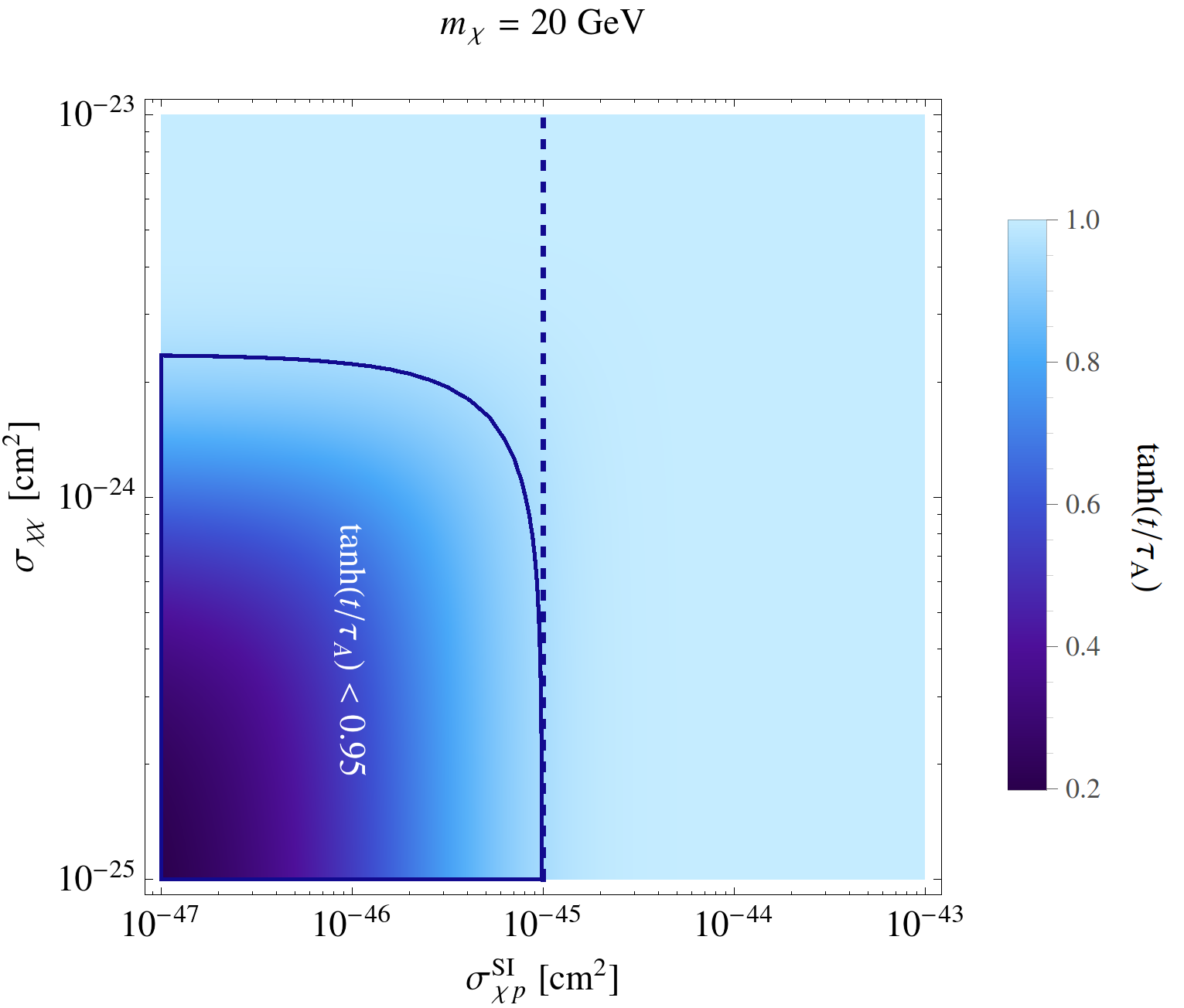}
\par\end{centering}
\caption{\label{fig:tanh_SI}The values of  $\tanh(t/\tau_{A})$ over $\sigma_{\chi p}^{\textrm{SI}}-\sigma_{\chi\chi}$
plane at the present day, $t=t_{\odot}$. The blue-circled area
is the non-equilibrium region for $N_{\chi}$. The vertical line at the right panel indicates the LUX bound, $\sigma_{\chi p}^{\textrm{SI}}\leq 10^{-45}~$cm$^2$, for $m_{\chi}=20$ GeV.   }
\end{figure}
The coefficients $C_{c,e,s}$ have been worked out in Refs.~\cite{Griest:1986yu,Gould:1987ju,Zentner2009}, which we adopt for our numerical studies. 
\begin{figure}[t]
\begin{centering}
\includegraphics[width=0.47\textwidth]{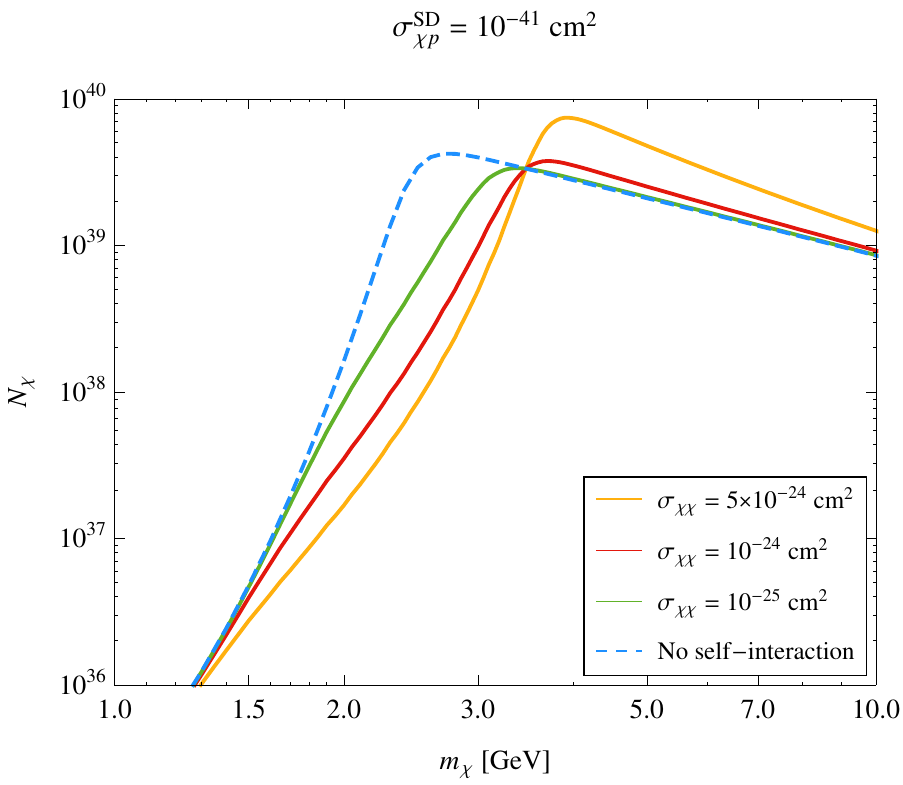} \includegraphics[width=0.47\textwidth]{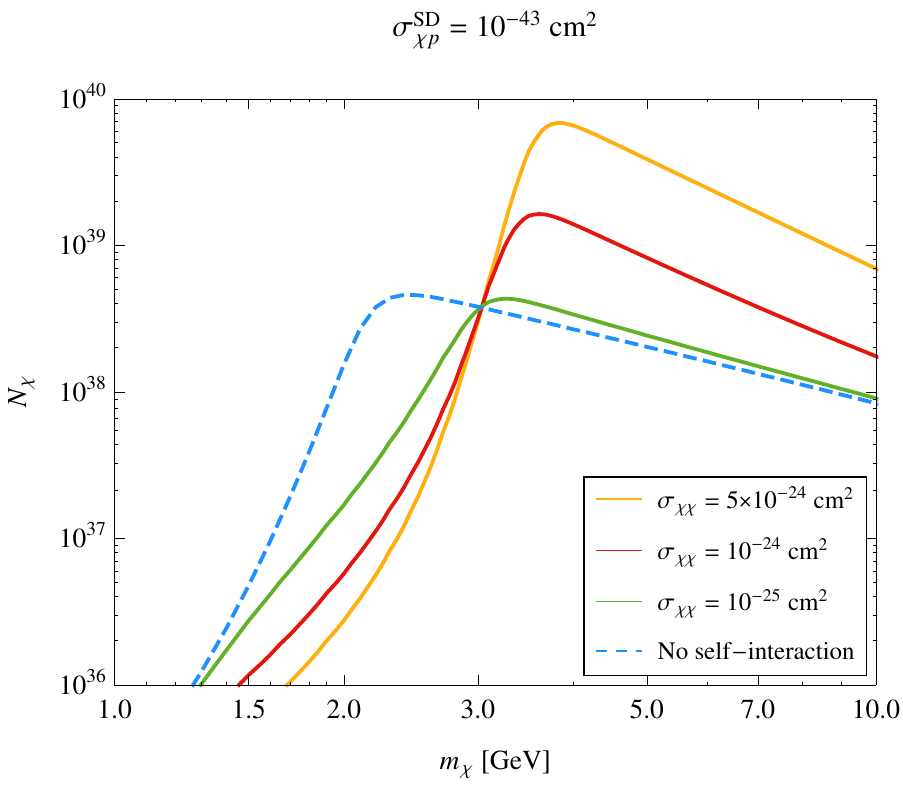} \includegraphics[width=0.47\textwidth]{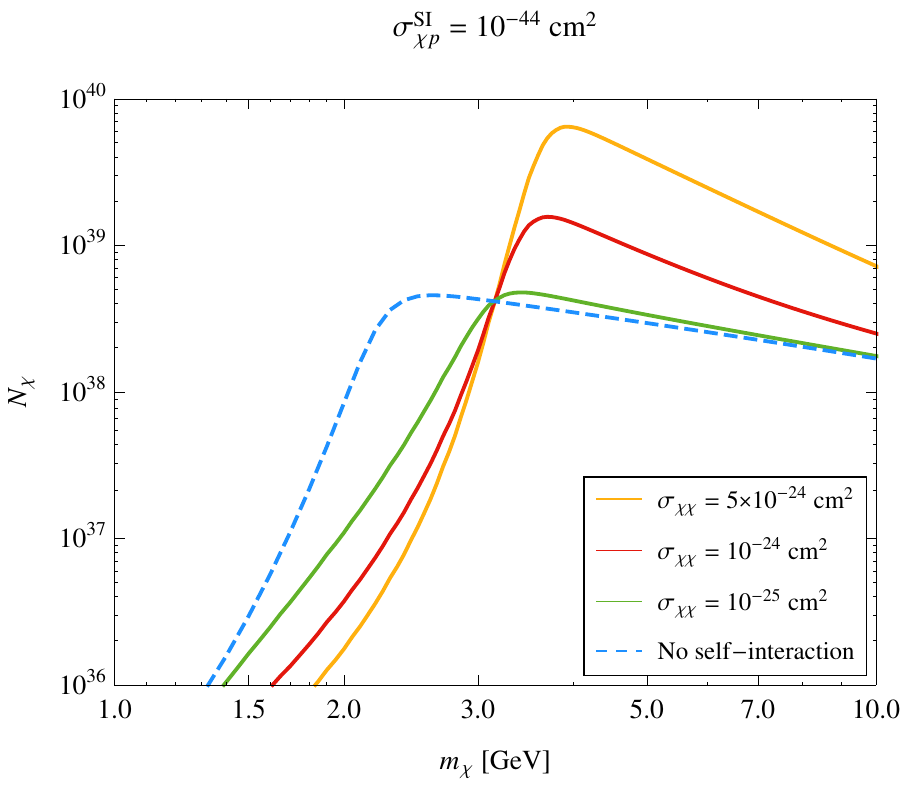} \includegraphics[width=0.47\textwidth]{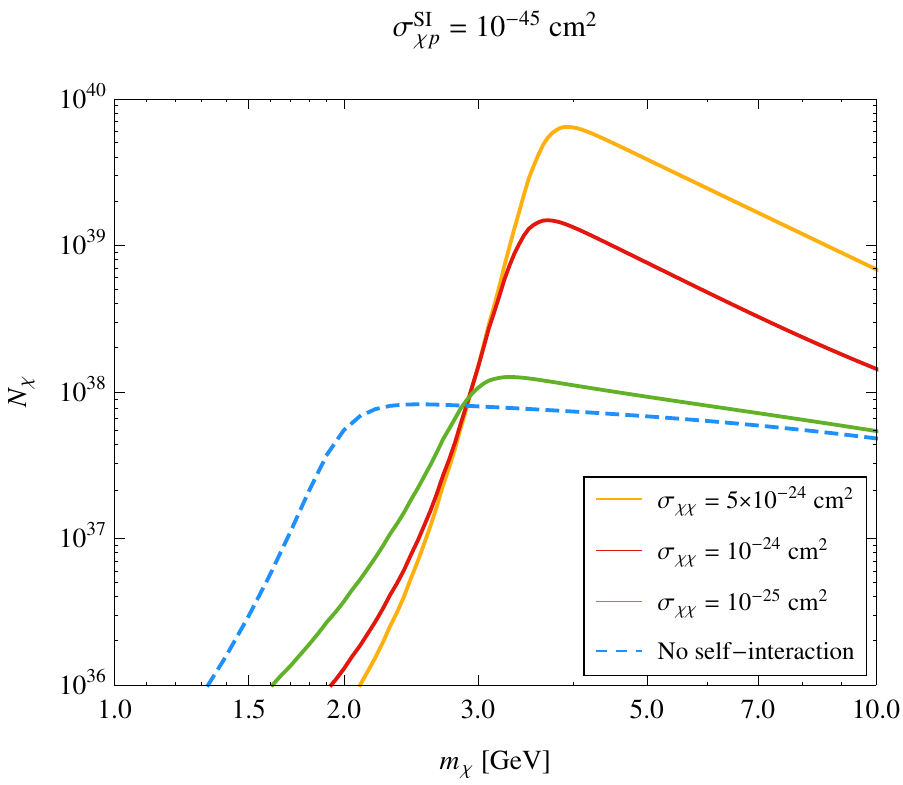}
\par\end{centering}
\caption{\label{fig:Nx_N0}The number of  DM particles trapped inside the Sun, $N_{\chi}$,  as a function of DM mass $m_{\chi}$ with and without DM self-interaction. Both SI and SD DM-nucleon couplings are considered. Different colors represent different values for $\sigma_{\chi\chi}$. The peak for each parameter set is the maximal DM number trapped inside the Sun, and the corresponding DM mass can be viewed as the evaporation mass scale since $N_{\chi}$ drops quickly for $m_{\chi}$ smaller than this mass scale.}
\end{figure}
The numerical result for $C_{se}$ is based on the analytic expression we have given in the Appendix.
We first identify the equilibrium region on  the $\sigma_{\chi p}$-$\sigma_{\chi\chi}$ plane. The SD and SI cases are presented in Fig.~\ref{fig:tanh_SD} and Fig.~\ref{fig:tanh_SI}, 
respectively with two benchmark DM masses  $m_{\chi}=5$~GeV and $20$~GeV. We have taken $10^{-45}\leq \sigma_{\chi p}^{\textrm{SD}}/{\textrm{cm}}^2\leq 10^{-41}$ and $10^{-47}\leq \sigma_{\chi p}^{\textrm{SI}}/{\textrm{cm}}^2\leq 10^{-43}$ for our studies.
The range of SD cross section is below those bounds set by direct detection experiments, COUPP~\cite{Behnke:2012ys} and Simple~\cite{Felizardo:2011uw}, and the indirect search by IceCube~\cite{Aartsen:2012kia} for $m_{\chi}\approx 20$  GeV. 
This range of SI cross section is below the direct detection bound set by LUX~\cite{LUX2013} at $m_{\chi}=5$~GeV.   For $m_{\chi}=20$
GeV, the LUX bound on $\sigma_{\chi p}^{\textrm{SI}}$ is $10^{-45}~$cm$^2$. We have indicated this bound on the right panel of Fig.~\ref{fig:tanh_SI}.
The dark areas represent those regions with $\tanh(t/\tau_{A})\lesssim1$ at the present
day whereas the light areas are the equilibrium regions.  

In Fig.~\ref{fig:Nx_N0}, we show the effect of DM self-interaction on the number of DM particles trapped inside the Sun. It is seen that $N_{\chi}$ can be significantly enhanced for sufficiently large $\sigma_{\chi\chi}$.  The $N_{\chi}$ peak for each parameter set is the maximal DM number trapped inside the Sun, and the corresponding DM mass can be viewed as the evaporation mass scale because $N_{\chi}$ drops quickly for $m_{\chi}$ smaller than this mass scale. We observe that the inclusion of DM self-interaction tends to lift the evaporation mass by about $1$~GeV. This is because the evaporation due to the captured DM-DM self-interaction comes to operate.
\begin{figure}[t]
\begin{centering}
\includegraphics[width=0.49\textwidth]{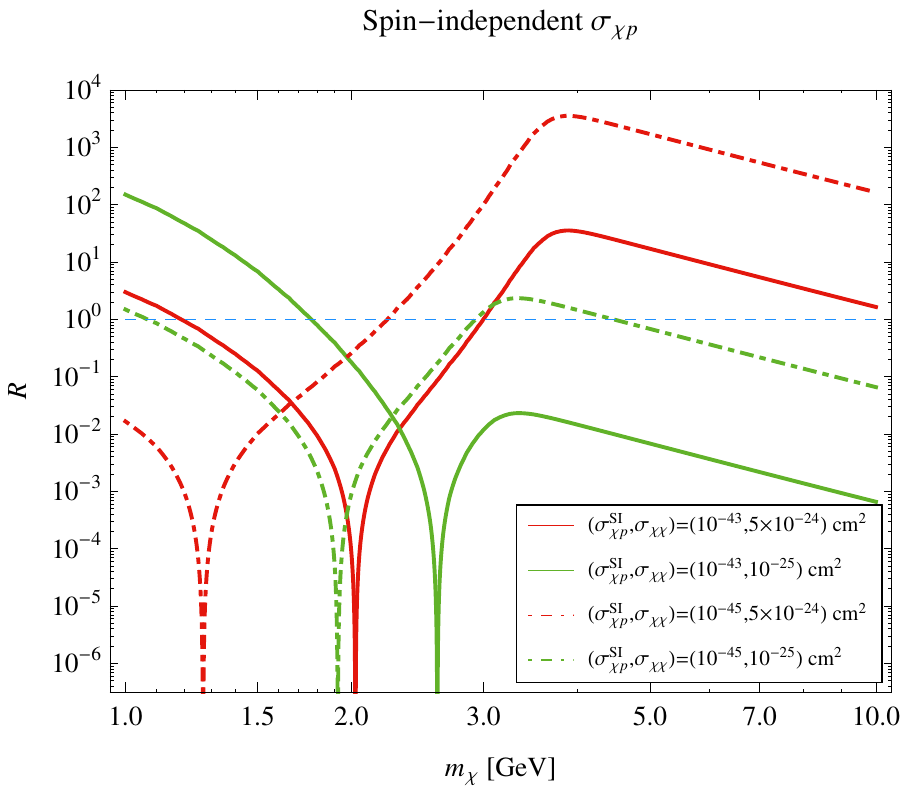}\includegraphics[width=0.49\textwidth]{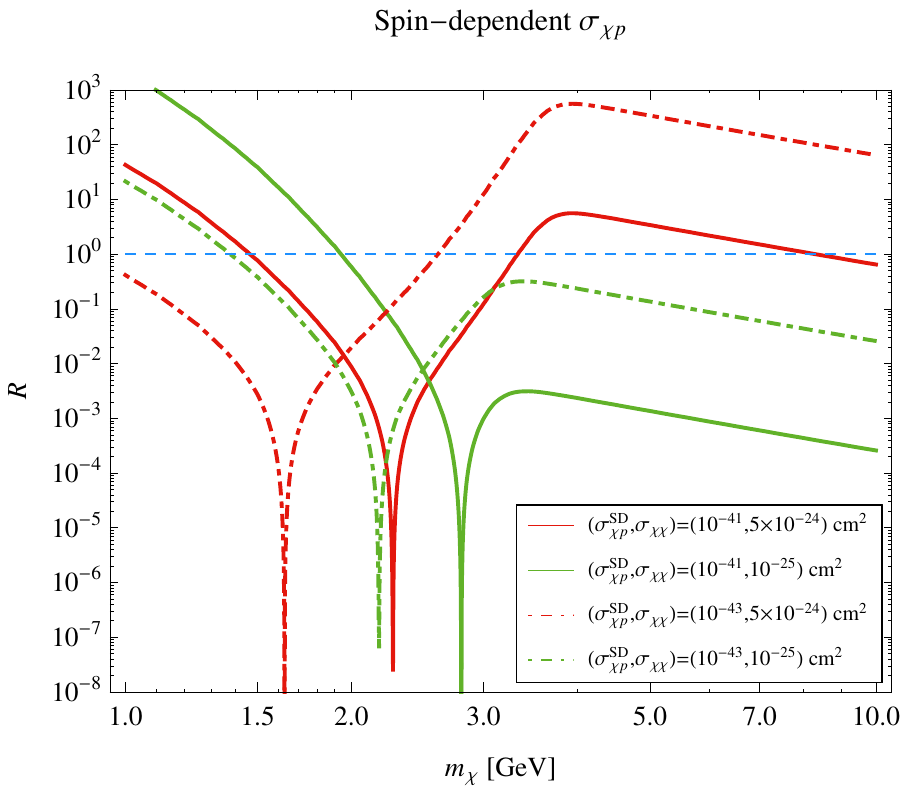}
\par\end{centering}
\caption{\label{fig:Rs_1D}Ratio $R$ versus DM mass $m_{\chi}$. The dip occurs when DM self-interaction and evaporation effects cancel each other, $C_{e} \approx C_{s}$. }
\end{figure}
\begin{figure}[t]
\begin{centering}
\includegraphics[width=0.49\textwidth]{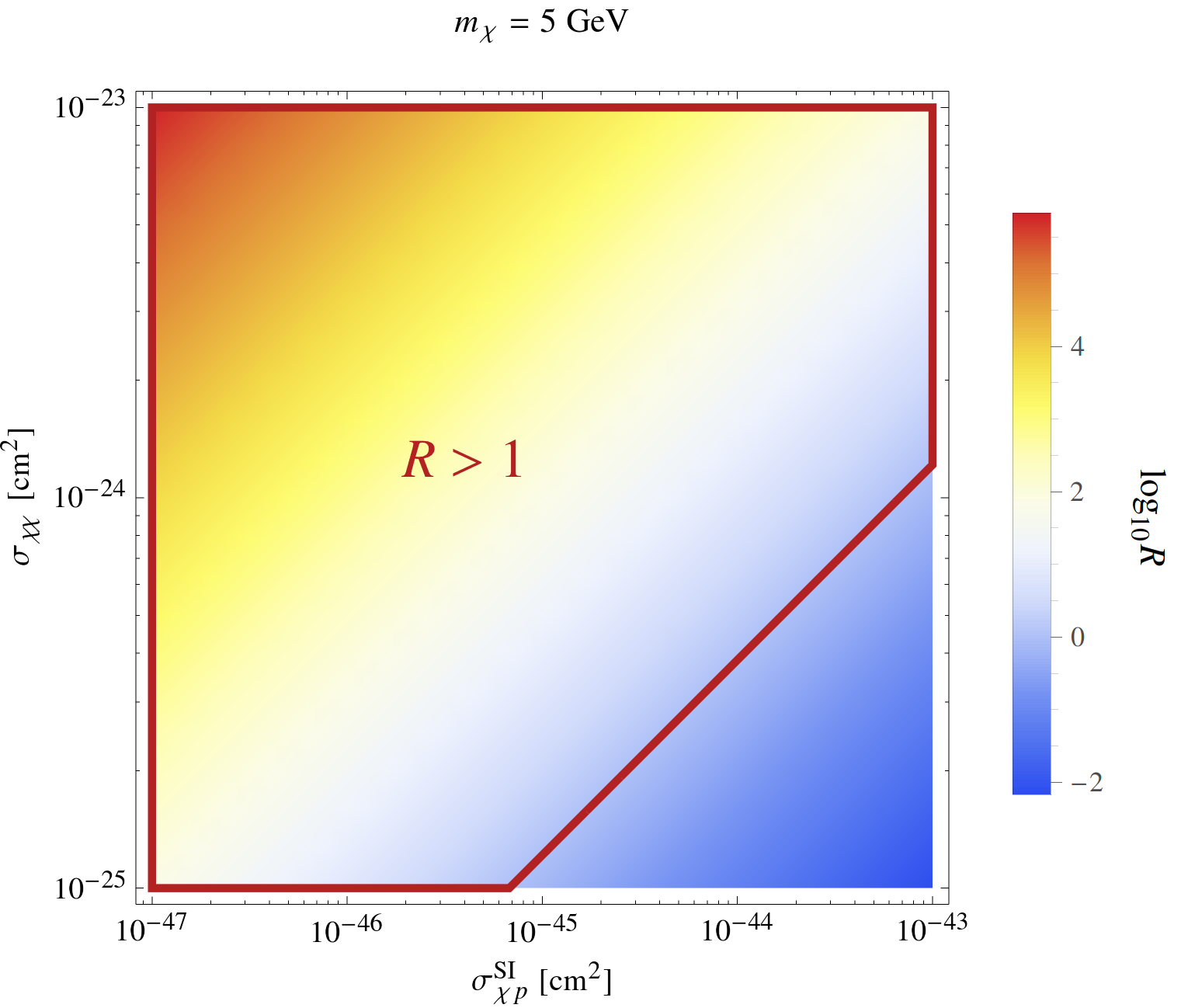}\includegraphics[width=0.49\textwidth]{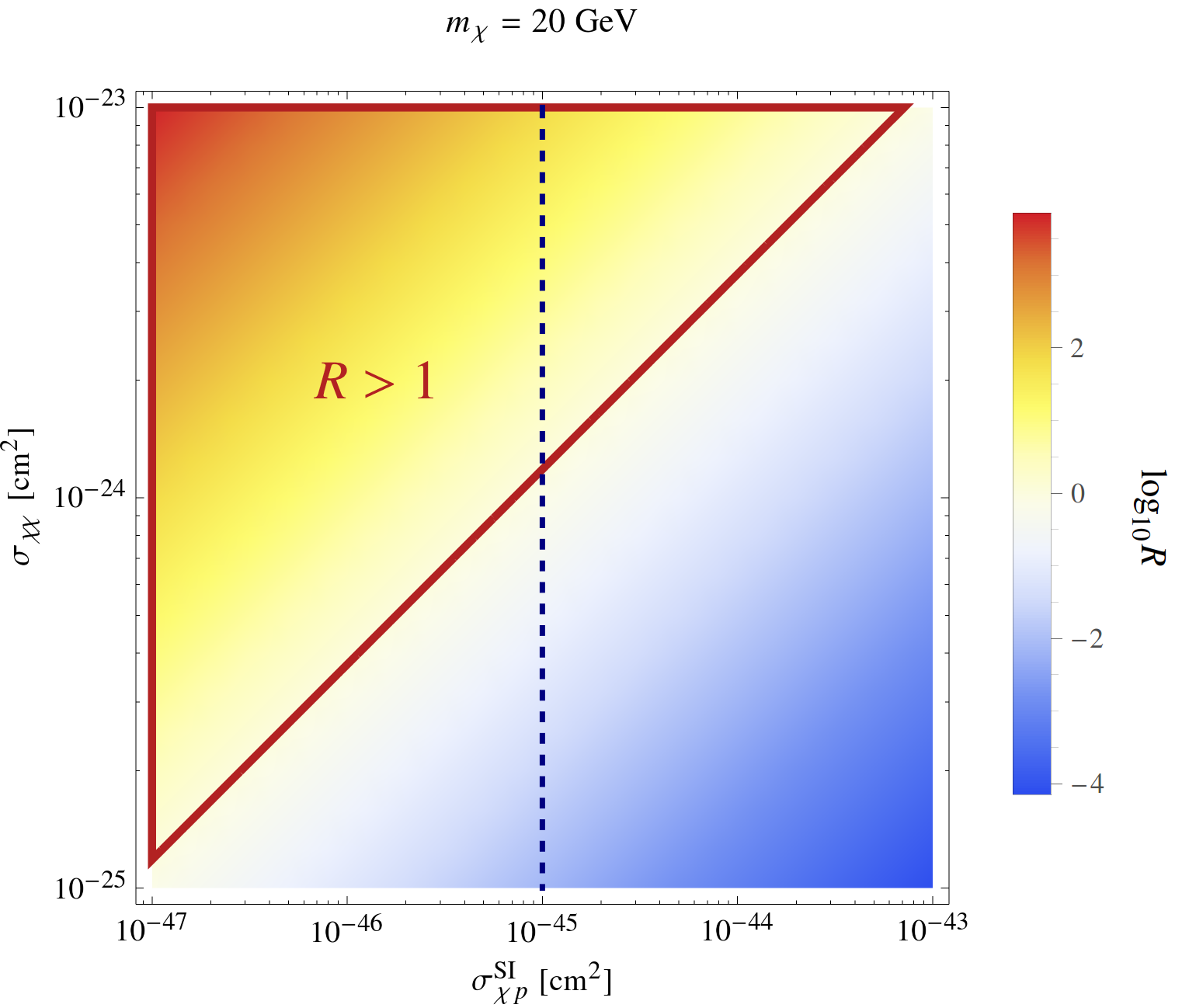}
\par\end{centering}
\begin{centering}
\includegraphics[width=0.49\textwidth]{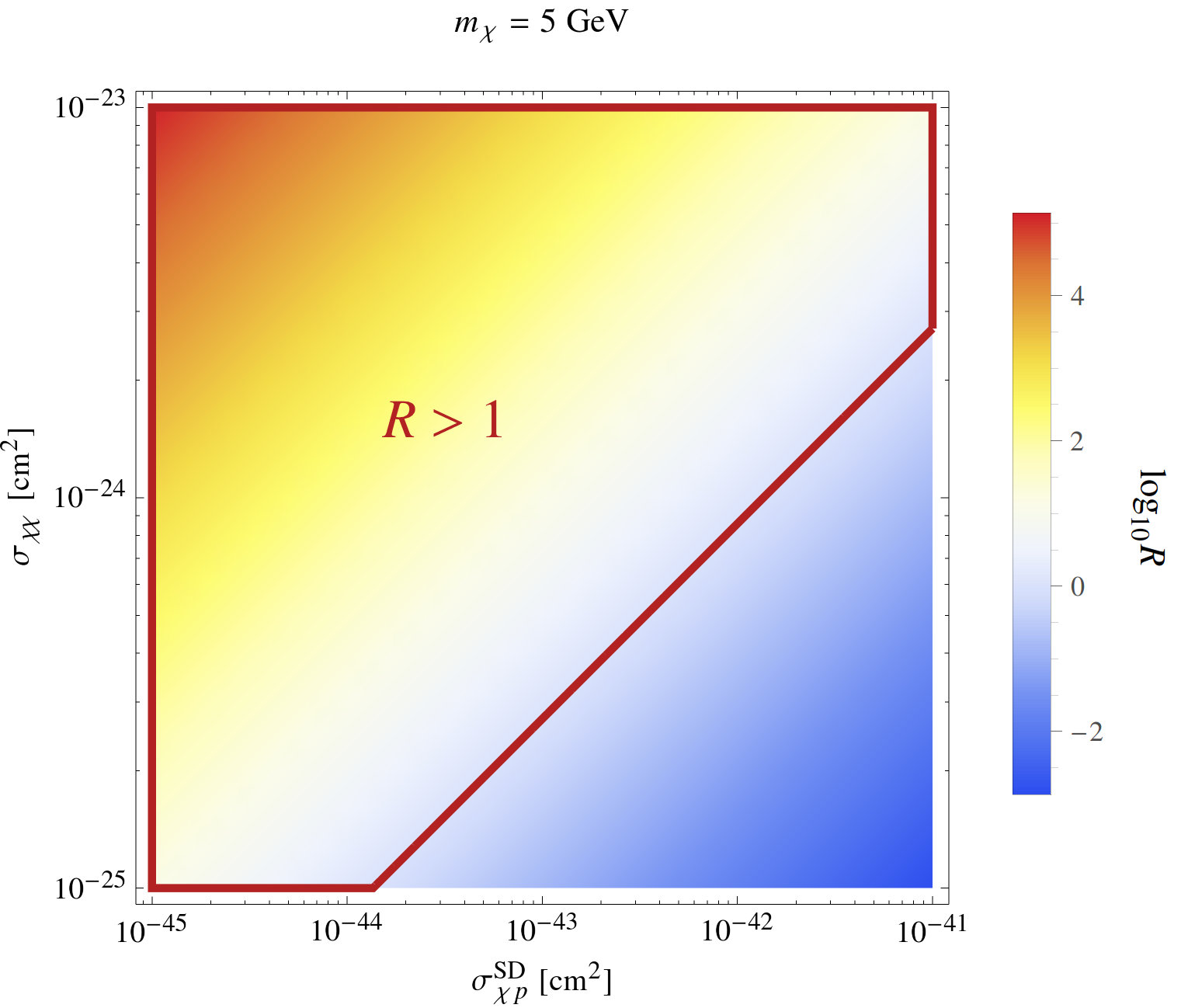}\includegraphics[width=0.49\textwidth]{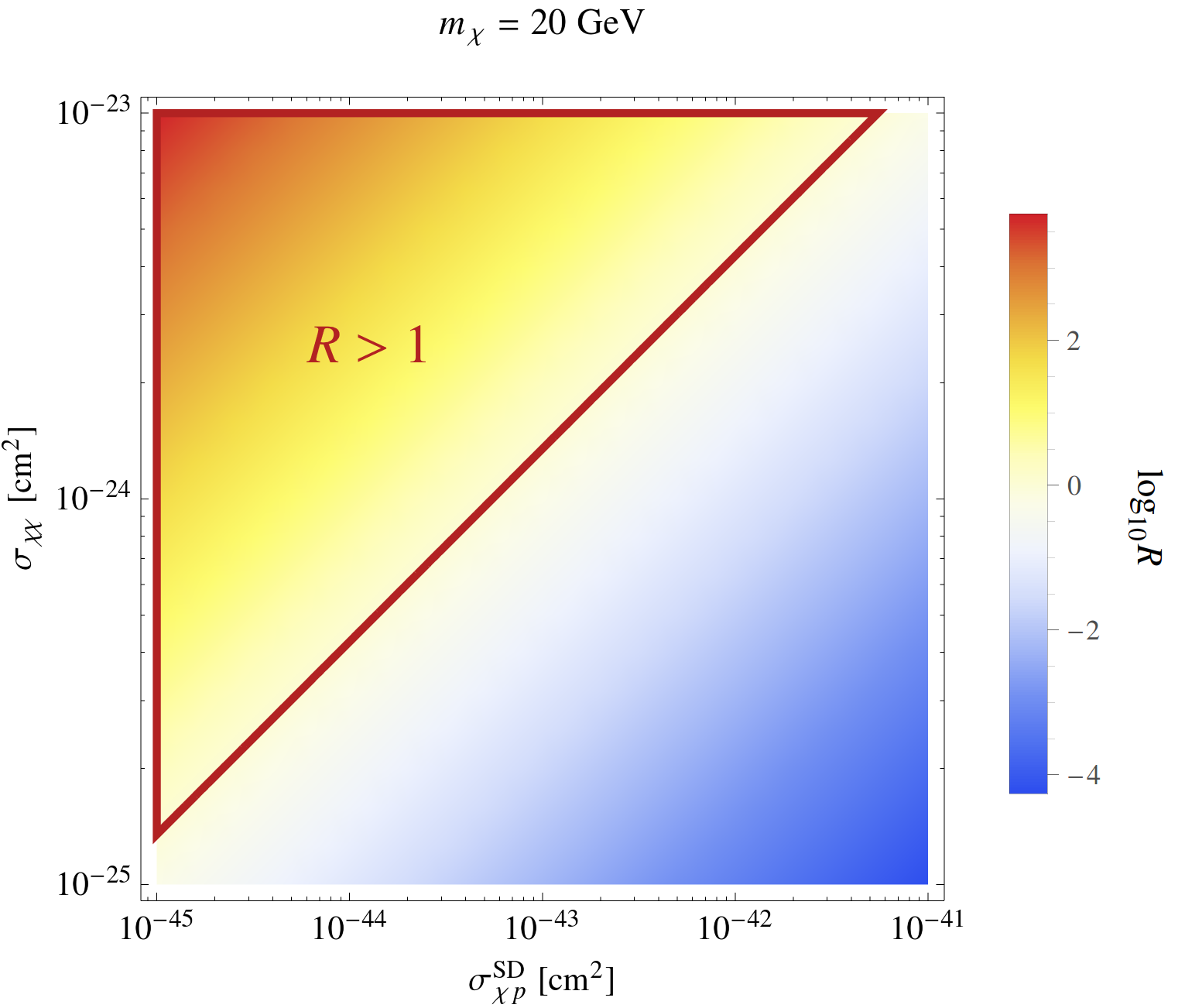}
\par\end{centering}
\caption{\label{fig:Rs_2D}Ratio $R$ over the $\sigma_{\chi p}-\sigma_{\chi\chi}$
plane. The upper panel is for SI interaction and the lower panel is for
SD interaction. The red-circled region is for $R>1$.}
\end{figure}

To quantify the effects of DM self-interaction and evaporation (the one by $C_e$) on $N_{\chi}$, it is useful to define a dimensionless parameter $R$ by
\begin{eqnarray}
R\equiv \frac{(C_{s}-C_{e})^{2}}{C_{c}(C_{a}+C_{se})},
\end{eqnarray}
such that
\begin{eqnarray}
N_{\chi,\textrm{eq}}=\sqrt{\frac{C_{c}}{C_{a}+C_{se}}}\left(\pm \sqrt{\frac{R}{4}}+\sqrt{\frac{R}{4}+1}\right),
\end{eqnarray}
and
\begin{eqnarray}
\Gamma_A=\frac{1}{2}\frac{C_{c}C_{a}}{C_{a}+C_{se}}\left(\pm \sqrt{\frac{R}{4}}+\sqrt{\frac{R}{4}+1}\right)^2,
\label{GaRse}
\end{eqnarray}
where one takes the positive sign for $C_s>C_e$ and the negative sign for $C_e>C_s$. The expression for $N_{\chi}$ in the non-equilibrium case can also be simplified in a similar way. 
Comparing $C_s$ with $C_e$, DM self-interaction dominates in the high mass region whereas the evaporation process takes over in the low mass region. 
Here we have taken into account both effects and found that the transition between two effects occurs at ${\cal O}(1)$ GeV DM mass. 
In Fig.~\ref{fig:Rs_1D}, we show the behavior of $R$ as a function of $m_{\chi}$. The dip of $R$ for each parameter set represents the narrow mass range where $C_{s} \approx C_{e}$. On the right side of the dip, $C_s$ dominates over $C_e$ while the reverse is true on the left side of the dip. 

It is seen that $R$ in the evaporation dominant region is growing up in small $m_{\chi}$  
since the velocity of final state $\chi$ after the collision can easily be larger than the escape velocity from the Sun in this case.  
The parameter space for $R > 1$ over the $\sigma_{\chi p}-\sigma_{\chi\chi}$ plane is shown in Fig.~\ref{fig:Rs_2D} for $m_{\chi} = 5$ and $20$~GeV. 
Since $C_e<C_s$ for the above chosen $m_{\chi}$, hence $R > 1$ is the region where DM self-interaction is relevant. 
It has been seen that DM self-interaction not only enhances $N_{\chi}$ significantly for  $m_{\chi}< 10$ GeV but also affects the evaporation mass scale.    
Therefore in the next session we shall explore the possibility of probing DM self-interaction for $m_{\chi}< 10$ GeV by IceCube-PINGU detector.

\section{Probing DM self-interaction at IceCube-PINGU}
\begin{figure}[t]
\begin{centering}
\includegraphics[width=0.49\textwidth]{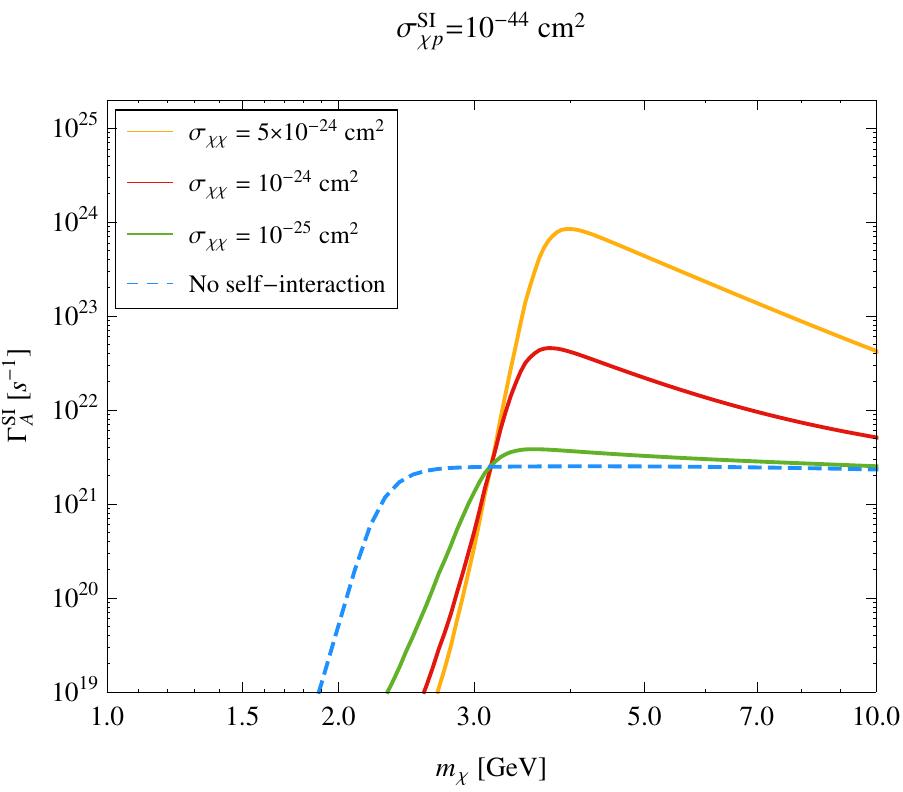}\includegraphics[width=0.49\textwidth]{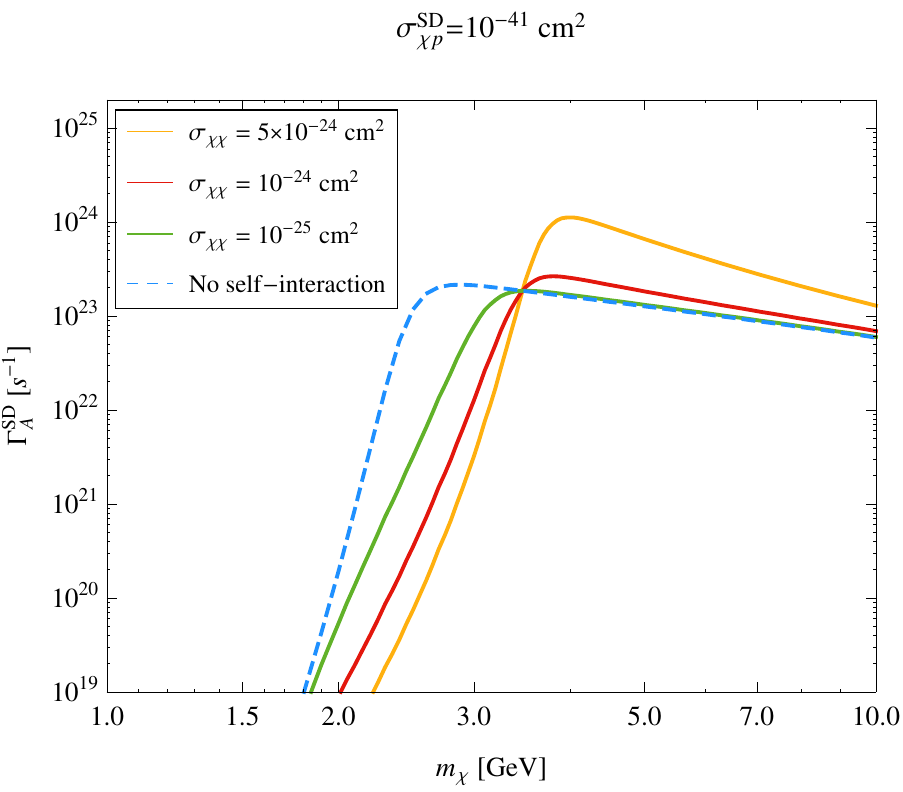}
\par\end{centering}
\caption{\label{fig:WIMP_number_ann}The annihilation rate $\Gamma_{A}$
of the captured DM inside the Sun. The left panel assumes DM-nuclei scattering is dominated by SI interaction while the right panel assumes such scattering is
dominated by SD interaction.}
\end{figure}
The annihilation rate of the captured DM in the Sun is given by Eq.~(\ref{annihilation}).
It is worth mentioning that in the absence of both evaporation (the one due to $C_e$) and self-interaction, the annihilation rate  $\Gamma_{A}$ with an equilibrium $N_{\chi}$ is
\begin{equation}
\Gamma_{A}=\frac{1}{2}C_{a}\times\frac{C_{c}}{C_{a}}=\frac{C_{c}}{2},
\end{equation}
which only depends on the capture rate $C_{c}$. However, with the presence of either $C_e$ or self-interaction, $\Gamma_{A}$ depends on other coefficients as well even $N_{\chi}$ has reached to the equilibrium. We plot $\Gamma_{A}$ as a function of $m_{\chi}$ with and without self-interaction in Fig.~\ref{fig:WIMP_number_ann}. 

To probe DM self-interaction for small $m_{\chi}$, we consider DM annihilation channels, 
$\chi\chi \rightarrow \tau^{+}\tau^{-}$ and $\nu\bar{\nu}$, for producing neutrino final states to be detected by IceCube-PINGU~\cite{Aartsen2014}.  The neutrino differential flux of flavor $i$, $\Phi_{\nu_{i}}$, from $\chi\chi\rightarrow f\bar{f}$
can be expressed as
\begin{equation}
\frac{d\Phi_{\nu_{i}}}{dE_{\nu_{i}}}=P_{\nu_j\to \nu_i}(E_{\nu})\frac{\Gamma_{A}}{4\pi R_{\odot}^{2}}\sum_{f}B_{f}\left(\frac{dN_{\nu_{j}}}{dE_{\nu_{j}}}\right)_{f}\label{eq:neutrino_flux}
\end{equation}
where $R_{\odot}$ is the distance between the neutrino source and
the detector, $P_{\nu_j\to \nu_i}(E_{\nu})$ is the neutrino oscillation
probability during the propagation, $B_{f}$ is the branching ratio
corresponding to the channel $\chi\chi\rightarrow f\bar{f}$ , $dN_{\nu}/dE_{\nu}$
is the neutrino spectrum at the source, and $\Gamma_{A}$ is the DM annihilation
rate in the Sun. To compute $dN_{\nu}/dE_{\nu}$, we employed \texttt{WimpSim}
\cite{Blennow:2007tw} with a total of 50,000 Monte-Carlo generated
events.

The neutrino event rate in the detector is given by
\begin{equation}
N_{\nu}=\int_{E_{\textrm{th}}}^{m_{\chi}}\frac{d\Phi_{\nu}}{ dE_{\nu}}A_{\nu}(E_{\nu})dE_{\nu}d\Omega\label{eq:nu_event}
\end{equation}
where $E_{\textrm{th}}$ is the detector threshold energy, $d\Phi_{\nu}/dE_{\nu}$
is the neutrino flux from DM annihilation, $A_{\nu}$ is the detector
effective area, and $\Omega$ is the solid angle. We study both muon track events and cascade events induced by neutrinos. The PINGU module will be implanted inside the IceCube in the near future~\cite{Aartsen2014} and can be used to probe neutrino energy down to ${\cal O}(1)$~GeVs. We take ice as the detector medium, so that the IceCube-PINGU neutrino effective area is expressed as 
\begin{equation}
A_{\textrm{eff}}^{\nu}(E_{\nu})=V_{\textrm{eff}}\frac{N_{A}}{M_{\textrm{ice}}}(n_{p}\sigma_{\nu p}(E_{\nu})+n_{n}\sigma_{\nu n}(E_{\nu})),
\end{equation}
where $V_{\textrm{eff}}$ is the IceCube-PINGU effective volume, $N_{A}$ is the Avogadro constant,
$M_{\textrm{ice}}$ is the mass of ice per mole, $n_{p,n}$ is the
number of proton/neutron of an ice molecule and $\sigma_{\nu p,n}$
is the neutrino-proton/neutron cross section which can be approximated by~\cite{Gandhi:1995tf}
\addtocounter{equation}{0}\begin{subequations}
\begin{align}
\frac{\sigma_{\nu N}(E_{\nu})}{E_{\nu}} & =6.66\times10^{-3}\textrm{ pb}\cdot\textrm{GeV}^{-1},\\
\frac{\sigma_{\bar{\nu}N}(E_{\nu})}{E_{\bar{\nu}}} & =3.25\times10^{-3}\textrm{ pb}\cdot\textrm{GeV}^{-1},
\end{align}
\end{subequations}
for $1\textrm{ GeV}\leq E_{\nu}\leq10\textrm{ GeV}$.
As the neutrinos propagate from the source to the detector, they encounter high-density
medium in the Sun, the vacuum in space, and the Earth medium. The matter
effect to the neutrino oscillation has been considered in $P_{\nu_j\to \nu_i}$ in Eq.~(\ref{eq:neutrino_flux}).

The atmospheric background event rate can also be calculated by Eq.~(\ref{eq:nu_event})
with $d\Phi_{\nu}/dE_{\nu}$ replaced by the atmospheric neutrino flux. Hence
\begin{equation}
N_{\textrm{atm}}=\int_{E_{\textrm{th}}}^{E_{\textrm{max}}}\frac{d\Phi_{\nu}^{\textrm{atm}}}{dE_{\nu}}A_{\nu}(E_{\nu})dE_{\nu}d\Omega.\label{eq:atm_event}
\end{equation}
In our calculation, the atmospheric neutrino flux $d\Phi_{\nu}^{\textrm{atm}}/dE_{\nu}$
is taken from Ref.~\cite{Aartsen:2012uu,Honda:2006qj}.
We set $E_{\textrm{max}}=m_{\chi}$ in order to compare with the DM signal.

\begin{figure}[t]
\begin{centering}
\includegraphics[width=0.45\textwidth]{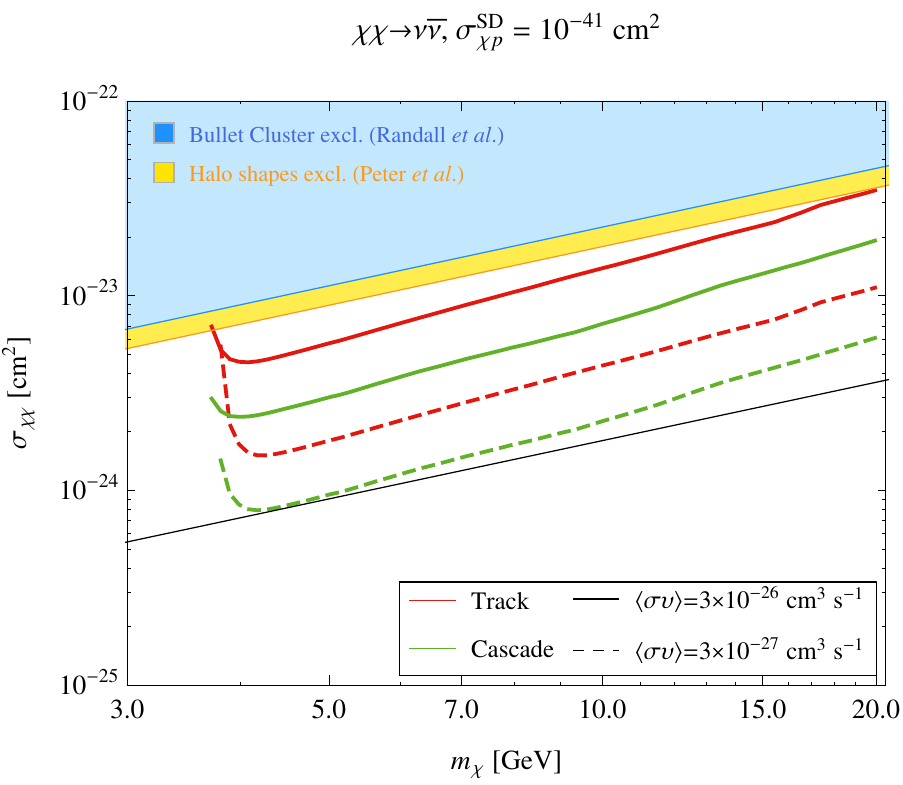} \includegraphics[width=0.45\textwidth]{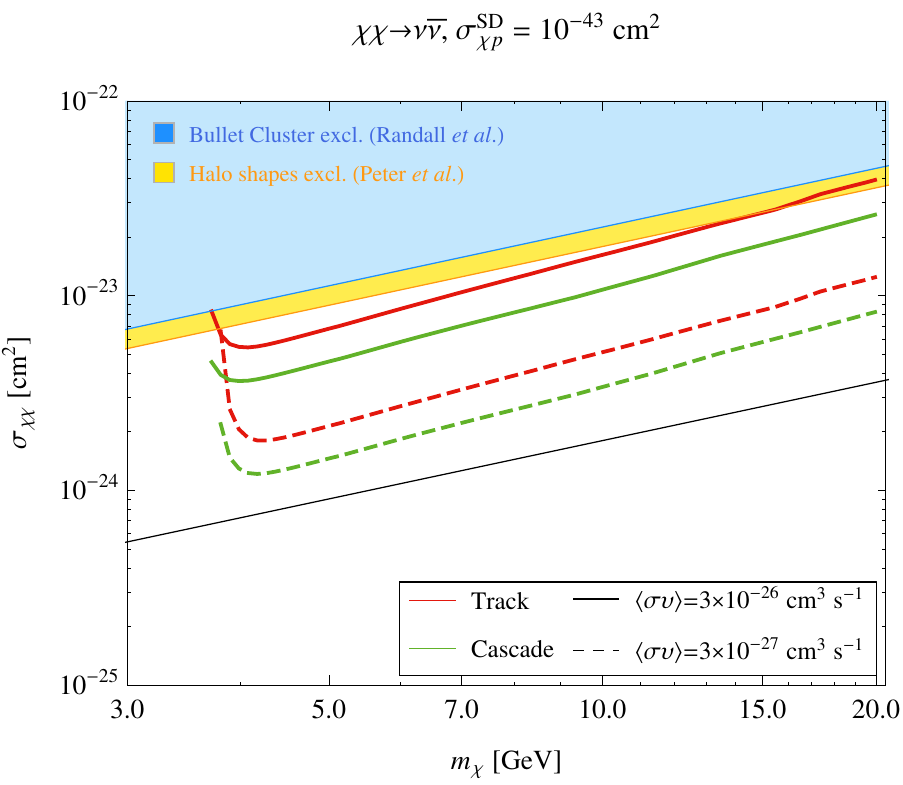}
\includegraphics[width=0.45\textwidth]{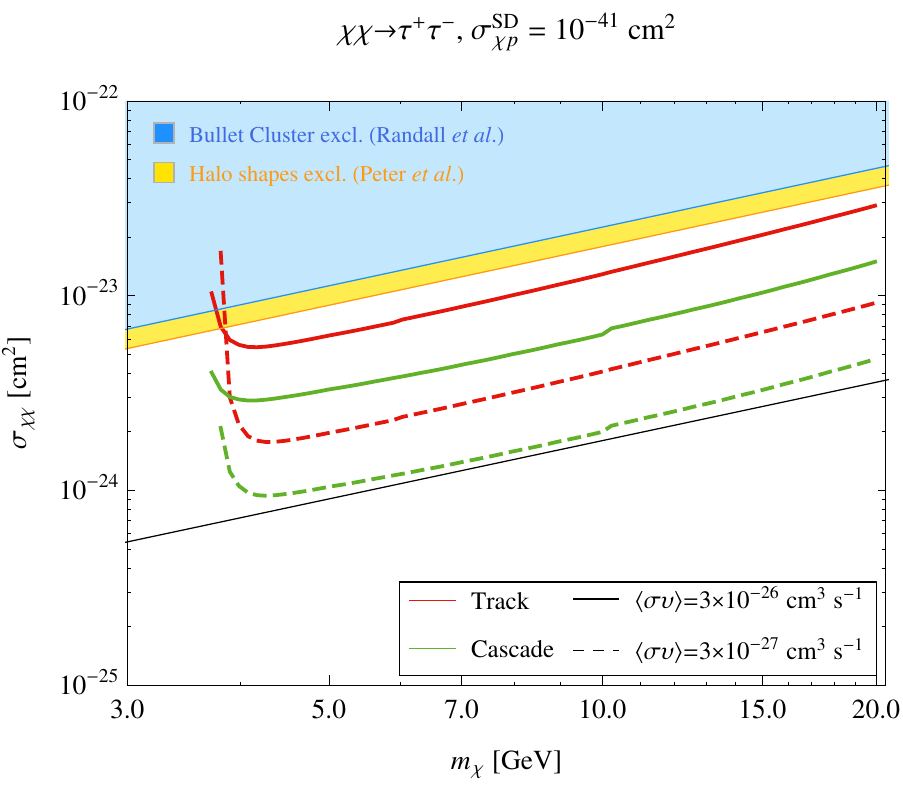}\includegraphics[width=0.45\textwidth]{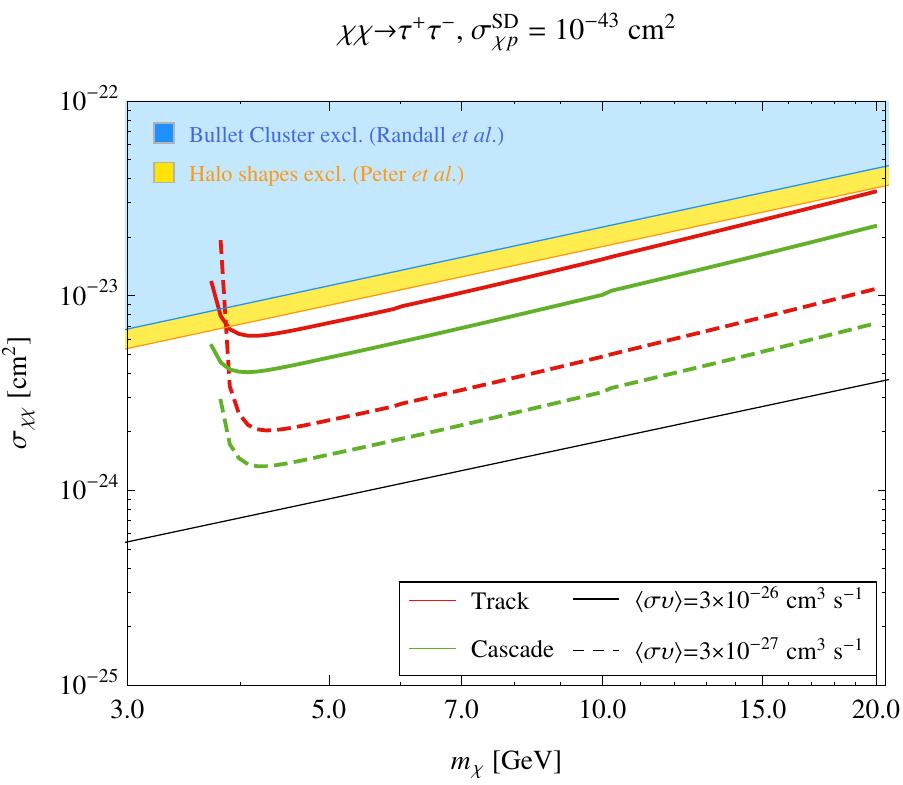}
\par\end{centering}
\caption{\label{fig:ConsSD}The IceCube-PINGU sensitivities to DM self-interaction cross section $\sigma_{\chi\chi}$ as a function of $m_{\chi}$. 
The DM-nucleus interaction inside the Sun is assumed to be dominated by SD interaction. }
\end{figure}

\begin{figure}[t]
\begin{centering}
\includegraphics[width=0.45\textwidth]{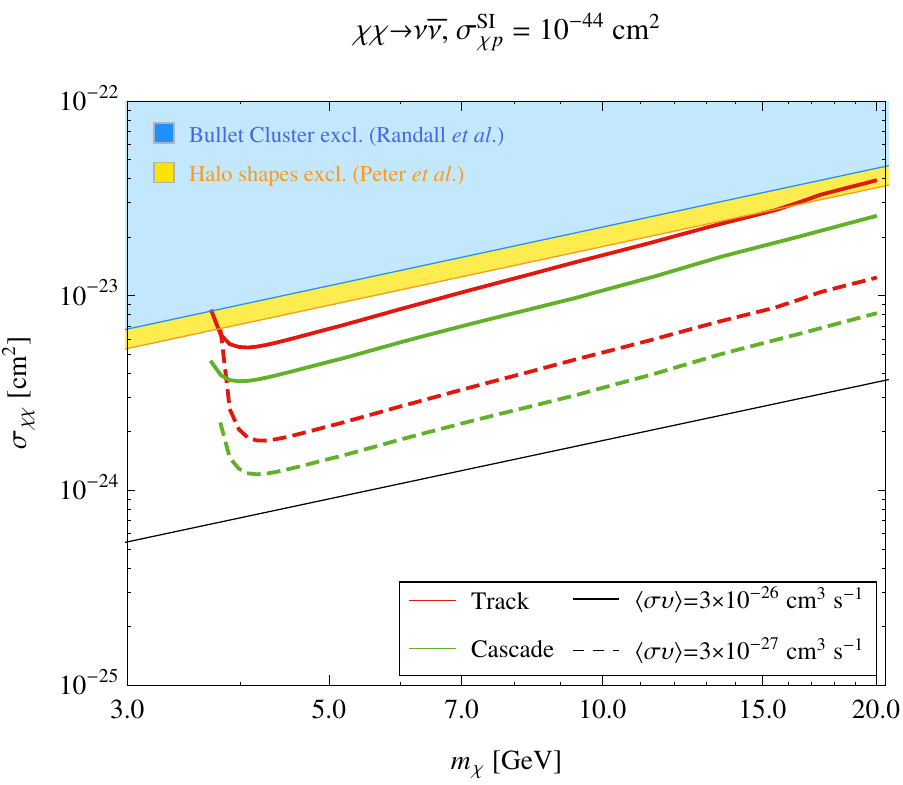} \includegraphics[width=0.45\textwidth]{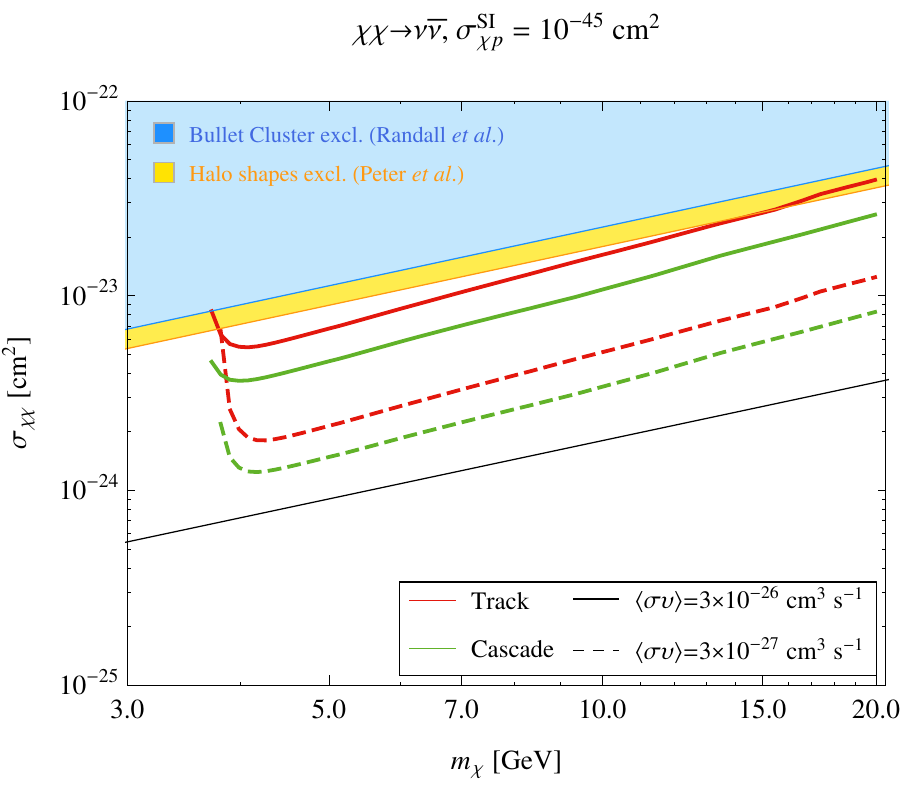}
\includegraphics[width=0.45\textwidth]{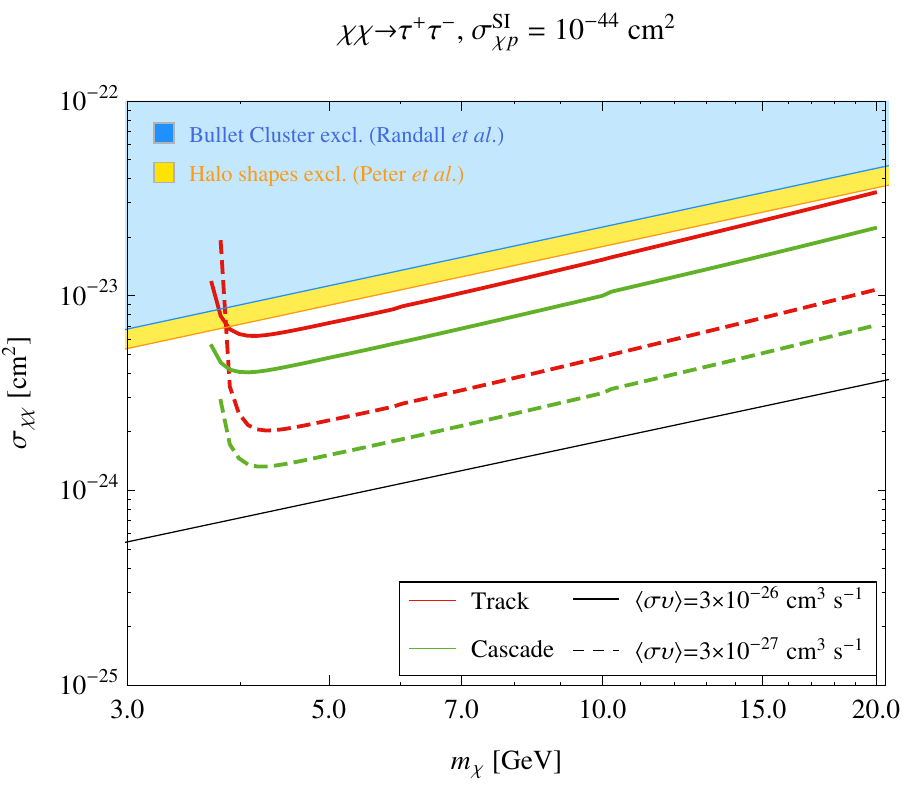}\includegraphics[width=0.45\textwidth]{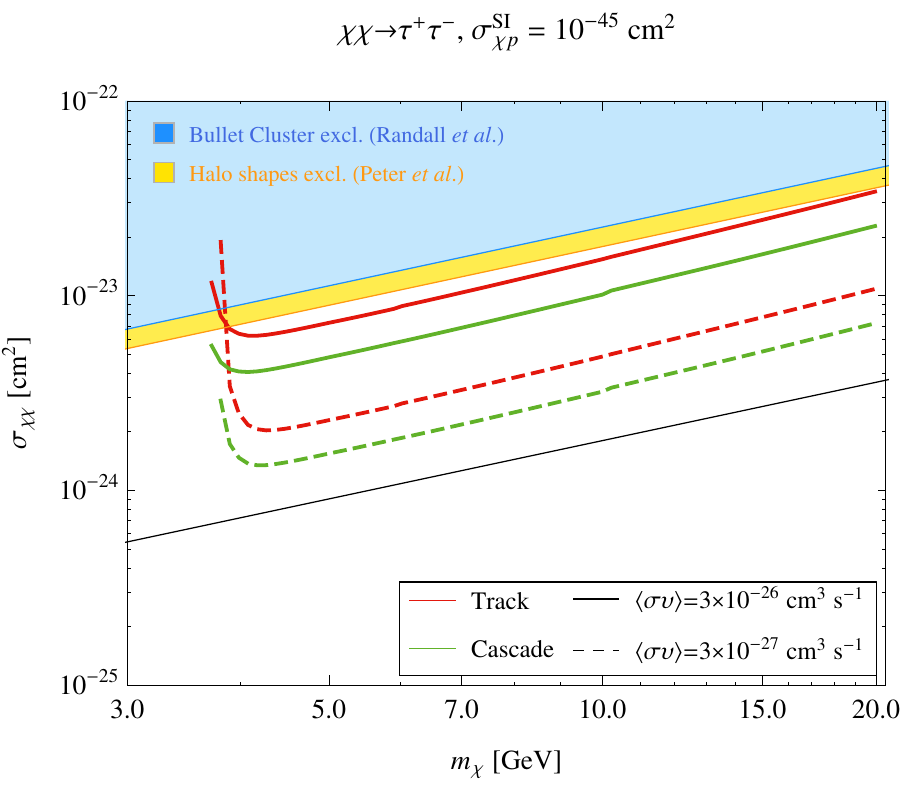}
\par\end{centering}
\caption{\label{fig:ConsSI}The IceCube-PINGU sensitivities to DM self-interaction cross section $\sigma_{\chi\chi}$ as a function of $m_{\chi}$. 
The DM-nucleus interaction inside the Sun is assumed to be dominated by SI interaction. 
}
\end{figure}

The angular resolution for IceCube-PINGU detector at $E_{\nu}=5$ GeV  is roughly $10^{\circ}$~\cite{Aartsen2014}. Hence we consider neutrino events arriving from the solid angle 
range $\Delta{\Omega}=2\pi(1-\cos\psi)$ surrounding the Sun with $\psi=10^{\circ}$.
We present the IceCube-PINGU sensitivity to $\sigma_{\chi\chi}$ in the DM mass region $3 ~{\rm GeV} < m_{\chi} < 20~{\rm GeV}$ for both SD and SI cases in Fig.~\ref{fig:ConsSD} and Fig.~\ref{fig:ConsSI}, respectively.  The sensitivities to $\sigma_{\chi\chi}$ are taken to be $2\sigma$ significance for 5 years of data taking. The shadow areas in the figures represent those parameter spaces disfavored by the Bullet Cluster and halo shape analyses. Below the black solid line, the DM self-interaction is too weak to resolve the core/cusp problem of the structure formation. Two benchmark values of thermal average cross section, $\langle \sigma v \rangle = 3\times10^{-26}~{\rm cm^3s^{-1}}~{\rm and}~\langle \sigma v \rangle =3\times 10^{-27}~{\rm cm^3s^{-1}}$ are used for our studies. We note that the latter value for $\langle \sigma v \rangle$ does not contradict with the relic density, since 
DM annihilation inside the Sun occurs much later than the period of freeze-out.  

We take  $\sigma_{\chi p}^{\rm SD} = 10^{-41}~{\rm cm^2}~{\rm and}~10^{-43}~{\rm cm^2}$ for SD interaction, and take  $\sigma_{\chi p}^{\rm SI} = 10^{-44}~{\rm cm^2}~{\rm and}~10^{-45}~{\rm cm^2}$ for SI interaction. We stress that $\sigma_{\chi p}^{\rm SD} = 10^{-41}~{\rm cm^2}$ is below the
lowest value of IceCube bound $\sigma_{\chi p}^{\rm SD} \sim 10^{-40}~{\rm cm^2}$ at $m_{\chi}\sim 300$ GeV~\cite{Aartsen:2012kia}.  For SI interaction,  
$\sigma_{\chi p}^{\rm SI} = 10^{-44}~{\rm cm^2}$ is below the LUX bound for $m_{\chi}< 8$ GeV, while $\sigma_{\chi p}^{\rm SI} = 10^{-45}~{\rm cm^2}$ is below the LUX bound
 for $m_{\chi}< 20$~GeV~\cite{LUX2013}.
We find that cascade events provide better sensitivities to DM self-interaction than track events do in all cases. One can also see that the sensitivity to $\sigma_{\chi\chi}$ becomes better for smaller annihilation cross section $\langle \sigma v \rangle$ for a fixed $\sigma_{\chi p}$, as noted in earlier works~\cite{Zentner2009, Albuquerque:2013xna} which neglect both $C_e$ and $C_{se}$. This is evident from Eq.~(\ref{GaRse}) since $R$ increases as $C_a$ decreases. It is instructive to take the limit $R\gg 1$ such that $\Gamma_A\to (C_{c}C_{a})R/2(C_{a}+C_{se})$ for $C_s>C_e$. It is easily seen that $\Gamma_A$ is inversely proportional to $C_a$ (in the mass range that $C_{se}$ is negligible) and is independent of $C_c$. In other words, only $C_s$ and $C_a$ determine the annihilation rate (we are in the region that $C_e$ is suppressed as compared to $C_s$).
We also see that the sensitivity to $\sigma_{\chi\chi}$ does become significantly worse as $m_{\chi}\to 4$ GeV.  
This is the critical $m_{\chi}$ below which the DM evaporations from the Sun is important.

\section{Conclusion}
We have studied the time evolution of DM number trapped inside the Sun with DM self-interaction considered. We have focused on the low $m_{\chi}$ range 
so that our analysis includes evaporation effects due to both DM-nuclei and DM-DM scatterings. The parameter region for the trapped DM inside the Sun to reach the equilibrium state is presented. We also found that the inclusion of DM self-interaction can increase the number of trapped DM and raise the evaporation mass scale. The parameter space on $\sigma_{\chi \chi}-\sigma_{\chi p}^{\rm SD\, (SI)}$ plane for significant enhancement on trapped DM number ($R>1$) is identified.  The parameter space for $R>1$ becomes larger for smaller $m_{\chi}$. For $C_s< C_e$, the condition $R>1$ leads to the suppression of neutrino flux, since the first term on the right hand side of Eq.~(\ref{GaRse}) is negative. 
We have proposed to study $\sigma_{\chi\chi}$ with the future IceCube-PINGU detector where the energy threshold can be lowered down to $1$ GeV.   
We considered cascade and track events resulting from neutrino flux induced by DM annihilation channels $\chi\chi\to \nu \bar{\nu}$ and $\chi\chi\to \tau^+ \tau^-$ inside the Sun.  We found that cascade events always provide better sensitivity to $\sigma_{\chi\chi}$. The sensitivity to
$\sigma_{\chi\chi}$ is also improved with a smaller DM annihilation cross section $\langle \sigma v \rangle $.
\section*{Acknowledgments\label{sec:ackn}}
We thank S. Palomares-Ruiz for a very useful comment. CSC is supported by the National
Center for Theoretical Sciences, Taiwan; FFL, GLL, and YHL are supported by Ministry of Science and Technology, Taiwan under 
Grant No. 102-2112-M-009-017.

\appendix

\section{\label{sec:appendix}DM self-interaction induced evaporation}

The derivation of DM self-interaction induced evaporation is similar
to the usual nucleon induced evaporation~\cite{Gould:1987ju,Kappl:2011kz,Bernal:2012qh,Busoni:2013kaa}. One simply makes the parameter replacements
\[
m_{N}\rightarrow m_{\chi},\; T_{N}\rightarrow T_{\chi},
\]
where $T_{\chi}$ is the DM temperature inside the Earth. The DM velocity distribution is approximated by Maxwell-Boltzmann 
distribution given by 
\[
f_{\odot}(w)=\frac{4}{\sqrt{\pi}}\left(\frac{m_{\chi}}{2T_{\chi}}\right)^{3/2}n_{\chi}w^{2}\exp\left(-\frac{m_{\chi}w^{2}}{2T_{\chi}}\right),
\]
where $w$ is the DM velocity.
The calculation of  DM-DM scattering rate proceeds by choosing one of the DM as the incident particle and the other DM as one of the targets which satisfy Maxwell-Boltzmann 
distribution in their velocities.  We then sum over the incident states with Maxwell-Boltzmann distribution as well. Let the velocity of the incident DM be $w$ and the velocity of the faster DM in the final state as $v$, respectively. Since we are considering the DM evaporation due to 
their self interactions, we have  $v>w$. 

It is essential to take note on the symmetry factor for identical particle scattering ($\chi\chi$ scattering) as compared to
the DM-nucleus scattering studied before. We note that the faster DM with velocity $v$  can be either one of the DM particles in the final state. This generates an extra factor of $2$  relative to DM-nucleus scattering.  On the other hand,
due to identical particles in the initial state, a factor $1/2$ must be applied as we sum over initial states according to thermal distributions. 
Hence the DM-DM differential scattering rate with the velocity transition $w\to v$ can be inferred from DM-nucleus scattering with suitable parameter replacements, which is given by
\begin{equation}
R^{+}(w\rightarrow v)dv=\frac{2}{\sqrt{\pi}}n_{\chi}\sigma_{\chi\chi}\frac{v}{w}e^{-\kappa^{2}(v^{2}-w^{2})}\chi(\beta_{-},\beta_{+})dv,\label{eq:R+}
\end{equation}
where we have summed up the target DM  according to the thermal distribution $f_{\odot}$ described above,
\[
\beta_{\pm}=\pm\kappa w \quad {\rm with} \quad 
\kappa = \sqrt{\frac{m_{\chi}}{2T_{\chi}}},
\]
and
\[
\chi(a,b)\equiv\int_{a}^{b}due^{-u^{2}}=\frac{\sqrt{\pi}}{2}[{\rm erf}(b)-{\rm erf}(a)].
\]
Therefore the DM-DM scattering rate with $v$ greater than the escape velocity $v_{\rm{esc}}$ is given by the integral
\begin{equation}
\Omega_{v_{\rm{esc}}}^{+}(w)=\int_{v_{\rm{esc}}}^{\infty}R^{+}(w\rightarrow v^{\prime})dv^{\prime}.
\end{equation}
Carrying out the integral yields
\begin{equation}
\Omega_{v_{\rm{esc}}}^{+}(w)=\frac{2}{\sqrt{\pi}}\frac{n_{\chi}\sigma_{\chi\chi}}{w}\frac{T_{\chi}}{m_{\chi}}\exp\left[\frac{m_{\chi}(v_{\rm{esc}}^{2}-w^{2})}{2T_{\chi}}\right]\chi(\beta_{-},\beta_{+}).
\end{equation}
To calculate the evaporation rate per unit volume at the position $\vec{r}$, we should sum up all possible states of the incident DM as follows: 
\begin{equation}
\frac{dC_{se}}{dV}=\int_{0}^{v_{\rm{esc}}}f_{\odot}(w)\Omega_{v_{\rm{esc}}}^{+}(w)dw.\label{eq:dE/dV}
\end{equation}
 
The DM number density inside the Sun is determined
by
\[
n_{\chi}(r)=n_{0}\exp\left(-\frac{m_{\chi}\phi(r)}{T_{\chi}}\right),
\]
where $n_{0}$ is the density at the solar core, and $\phi(r)$ is the solar gravitational potential with respect
to the core so that
\[
\phi(r)=\int_{0}^{r}\frac{GM_{\odot}(r^{\prime})}{r^{\prime2}}dr^{\prime},
\]
with $G$ the Newton gravitational constant, $M_{\odot}(r)=4\pi\int_{0}^{r}r^{\prime2}\rho_{\odot}(r^{\prime})dr^{\prime}$
the solar mass enclosed within radius $r$, and $\rho_{\odot}(r)$
the solar density. 
Thus, the integration in Eq.~(\ref{eq:dE/dV}) can be performed
\begin{equation}
\frac{dC_{se}}{dV}=\frac{4}{\sqrt{\pi}}\sqrt{\frac{m_{\chi}}{2T_{\chi}}}\frac{n_{0}^{2}\sigma_{\chi\chi}}{m_{\chi}}\exp\left[-\frac{2m_{\chi}\phi(r)}{T_{\chi}}\right]
\exp\left[-\frac{E_{{\rm esc}}(r)}{T_{\chi}}\right]\tilde{K}(m_{\chi})
\end{equation}
where
\begin{equation}
\tilde{K}(m_{\chi})=\sqrt{\frac{E_{{\rm esc}}(r)T_{\chi}}{\pi}}\exp\left[-\frac{E_{{\rm esc}}(r)}{T_{\chi}}\right]+\left(E_{{\rm esc}}(r)-\frac{T_{\chi}}{2}\right){\rm erf}\left(\sqrt{\frac{E_{{\rm esc}}(r)}{T_{\chi}}}\right),
\end{equation}
with $E_{{\rm esc}}(r)$ the escape energy at the position $r$ defined by
\begin{equation}
E_{{\rm esc}}(r)=\frac{1}{2}m_{\chi}v_{\rm{esc}}^{2}(r).
\end{equation}
The escape velocity  $v_{\rm{esc}}(r)$ is related to the gravitational potential by $v_{\rm{esc}}(r)\equiv\sqrt{2[\phi(\infty)-\phi(r)]}$.
Finally, the self-interaction induced evaporation rate can be evaluated
through the following:
\begin{equation}
C_{se}=\frac{\int_{\odot}\frac{dC_{se}}{dV}d^{3}r}{\left(\int_{\odot}n_{\chi}(r)d^{3}r\right)^{2}}.
\end{equation}

\end{document}